\documentclass[journal]{IEEEtran}

\usepackage{amsfonts}

\usepackage{enumitem}
\usepackage{color}
\usepackage{graphicx}
\usepackage{booktabs,array}
\usepackage{multirow}
\usepackage{graphicx}
\usepackage{cite}

\usepackage[breaklinks, bookmarks = false]{hyperref}
\hypersetup{
  colorlinks   = true,
  urlcolor     = blue,
  linkcolor    = red,
  citecolor   = blue
}

\def\BibTeX{{\rm B\kern-.05em{\sc i\kern-.025em b}\kern-.08em
    T\kern-.1667em\lower.7ex\hbox{E}\kern-.125emX}}

\usepackage{caption}
\renewcommand{\figurename}{Fig.}
\renewcommand{\thefigure}{\arabic{figure}}

\begin{document}

\title{\Large \bf Revisiting Game-Theoretic Control in Socio-Technical Networks: Emerging Design Frameworks and Contemporary Applications}

\author{Quanyan Zhu,~\IEEEmembership{Senior Member,~IEEE}, and Tamer Ba\c{s}ar, ~\IEEEmembership{Life Fellow,~IEEE}\thanks{Quanyan Zhu is with New York University, New York, NY 11201, USA; E-mail: qz494@nyu.edu.
Tamer Ba\c{s}ar is with the Coordinated Science Laboratory, University of Illinois Urbana-Champaign, Urbana, IL 61801, USA; E-mail: basar1@illinois.edu.} }
\maketitle

\begin{abstract}
Socio-technical networks represent emerging cyber-physical infrastructures that are tightly interwoven with human networks. The coupling between human and technical networks presents significant challenges in managing, controlling, and securing these complex, interdependent systems. This paper investigates game-theoretic frameworks for the design and control of socio-technical networks, with a focus on critical applications such as misinformation management, infrastructure optimization, and resilience in socio-cyber-physical systems (SCPS). Core methodologies, including Stackelberg games, mechanism design, and dynamic game theory, are examined as powerful tools for modeling interactions in hierarchical, multi-agent environments. Key challenges addressed include mitigating human-driven vulnerabilities, managing large-scale system dynamics, and countering adversarial threats. By bridging individual agent behaviors with overarching system goals, this work illustrates how the integration of game theory and control theory can lead to robust, resilient, and adaptive socio-technical networks. This paper highlights the potential of these frameworks to dynamically align decentralized agent actions with system-wide objectives of stability, security, and efficiency.
\end{abstract}


\IEEEpeerreviewmaketitle

\ifCLASSOPTIONcaptionsoff
  \newpage
\fi

\section{Introduction}
\label{introduction}
Game theory addresses strategic interactions among decision-makers, often referred to as players or agents \cite{vonNeumann1944}. Each player has a distinct objective function—either a utility to maximize or a cost to minimize—which encapsulates their preferences across available alternatives. However, these preferences are interdependent, shaped by the choices made by other players, creating a need for game theory as a framework to model these strategic dynamics \cite{nash1950}.

In non-cooperative games, where players act independently, achieving an equilibrium is a central focus. The Nash equilibrium, a key concept developed by John Nash, represents a stable state where no player can unilaterally adjust their strategy for a better outcome \cite{nash1951}. This equilibrium concept enables analysts to identify stable states in competitive environments. When hierarchical decision-making exists, other solution concepts like the Stackelberg equilibrium are more suitable, especially in scenarios where leaders act first and followers respond. This framework is particularly applicable in control scenarios where independent agents must coordinate within shared constraints \cite{von2010market}.

The formalization of game theory is largely attributed to John von Neumann and Oskar Morgenstern’s \textit{Theory of Games and Economic Behavior} \cite{vonNeumann1944}, a pioneering text that established the field’s foundations and generated broad, interdisciplinary interest. Nash’s work on equilibrium expanded the field significantly, enabling stable solutions for non-cooperative games \cite{nash1951}. Additional advancements by Richard Bellman, who introduced dynamic programming \cite{bellman1957}, and Rufus Isaacs, known for differential games \cite{isaacs1965}, extended game theory to dynamic and multi-stage decision-making, integrating it more deeply into control and optimization contexts.

A defining moment for game theory was its integration into optimal control and decision processes, especially through the efforts of researchers at the RAND Corporation, including Bellman, Nash, and Isaacs. At RAND, these theorists developed models intersecting military strategy, economics, and control systems, catalyzing breakthroughs in multi-agent decision-making processes. Isaacs' differential games, for instance, established a framework for continuous-time strategic interactions' foundational for adversarial scenarios like pursuit-evasion games \cite{isaacs1965}. This interdisciplinary evolution of game theory alongside advances in optimal control provided the groundwork for its application in modern networked systems, where it remains central to understanding strategic interdependencies and designing resilient control mechanisms.

\subsection{Game Theory and Socio-Technical Systems}
Today’s interconnected systems—ranging from telecommunications and social networks to critical infrastructure—face unique challenges as they are populated by autonomous agents, each pursuing individual objectives, but interconnected through their actions and information exchanges. Game theory is a critical tool in such environments, especially where decentralized decision-making is needed. In multi-agent systems, every agent’s actions affect the network’s overall state, often creating complex dynamics that are difficult to predict and control. Game-theoretic analysis equips researchers and designers with the methods to anticipate these interactions, predict system behavior, and develop strategies that enhance stability, efficiency, and resilience across the network \cite{fudenberg1991}.

Moreover, these networks are often socio-technical systems, where human behavior directly influences their performance, efficiency, and resilience. Human decisions and interactions shape the functioning of many networked systems, such as transportation, energy grids, and public health infrastructure \cite{helbing2013,zimmerman2017conceptual,zimmerman2016promoting}. For example, transportation networks must account for the flexible and sometimes unpredictable nature of human routing decisions. In such settings, infrastructure planning cannot be isolated from human behavior. The well-known Braess paradox illustrates that adding roads to a network may lead to increased congestion, as drivers individually optimize their routes, often at the expense of overall efficiency \cite{braess1968}. Game-theoretic models help planners anticipate these outcomes, enabling the design of transportation networks that mitigate unintended consequences and improve flow \cite{roughgarden2005}.

\begin{figure}
    \centering
    \includegraphics[width=0.8\linewidth]{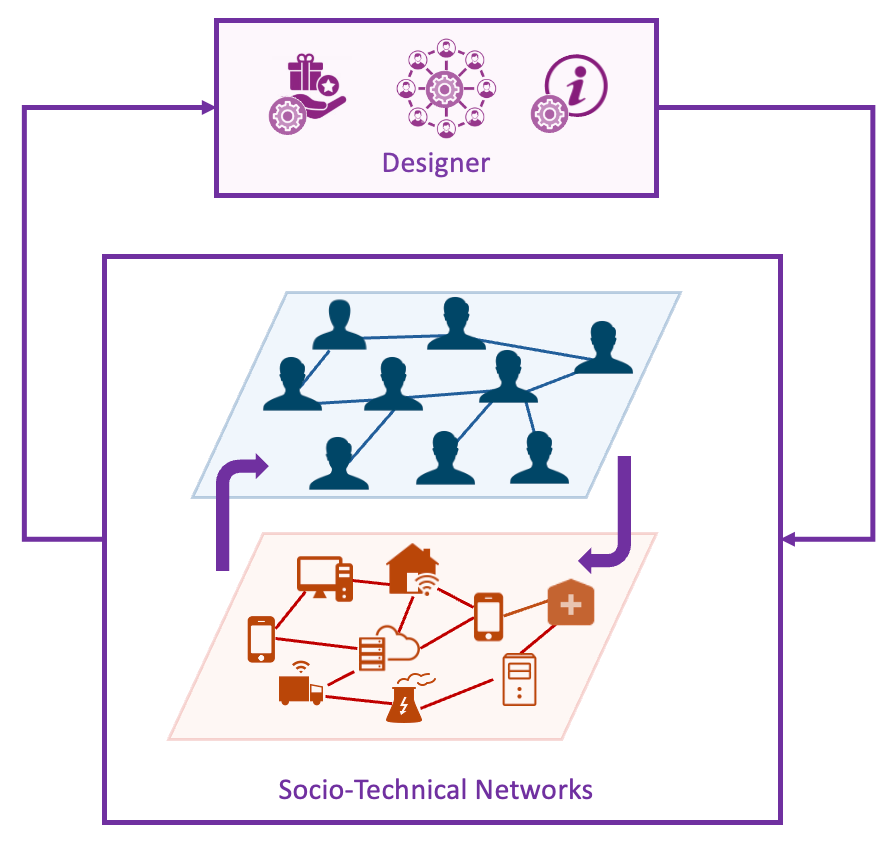}
    \caption{A Game-Theoretic Control Paradigm for Socio-Technical Systems: Socio-technical networks are composed of interconnected human and technical networks. Human agents interact both with one another and with technical infrastructures, including power grids, transportation systems, and cyber networks. The control of these networks can be achieved through strategic designs in information flow, network structure, and incentive mechanisms. Information design guides how agents access and process data, while network design shapes the connectivity and interaction pathways within the system. Incentive design, on the other hand, motivates desired behaviors by aligning agent actions with system-wide objectives, ensuring that human and technical interactions are coordinated to achieve resilience, efficiency, and security across the socio-technical network.}\label{overall}
\end{figure}

In smart energy systems, where energy prosumers (both consumers and producers) decide when to buy or sell energy, individual behaviors impact the grid’s supply-demand balance. To maintain grid stability, game-theoretic mechanisms can incentivize prosumers to make decisions that align personal economic interests with the system’s operational needs \cite{saad2012}. These control mechanisms foster a resilient and efficient energy network by balancing incentives for prosumers in ways that enhance resource allocation and minimize the risk of outages \cite{stoustrup2019smart}. Figure \ref{overall} illustrates the nature of the control of socio-technical systems. The technical system is coupled with the human networks, and the designer can influence the coupled system through different control paradigms, including information, incentives, and network structures.

Public health further exemplifies the socio-technical nature of modern systems, as seen during the COVID-19 pandemic. Individual choices, such as decisions about vaccination, mask-wearing, and social interactions, had substantial effects on the spread of the virus \cite{liu2022herd,PBB20,gubar2018optimal,hayel2017epidemic}. In such interconnected populations, the community’s health state depends on the aggregation of personal decisions. Game-theoretic design principles offer powerful tools for crafting behavioral incentives and nudges that guide individuals toward compliance with public health measures \cite{zhu2022preface}. By structuring these incentives effectively, game theory helps manage collective health outcomes, particularly during health crises, and underscores the importance of considering socio-technical dynamics in system design.

By integrating game-theoretic frameworks into socio-technical systems, designers gain the ability to understand and anticipate human-driven impacts on system dynamics. Game theory provides structured approaches for designing incentives, controlling information flows, and implementing adaptive mechanisms that foster desirable behaviors. These strategies are essential for ensuring that the interactions of autonomous agents—whether they are people, machines, or a mix of both—contribute positively to network performance, resilience, and societal benefit \cite{myerson1991}.

\subsection{Game-Theoretic Control Design}
Game theory offers not only a framework for modeling, performance evaluation, and risk assessment but also a robust design methodology for creating decentralized agents. A key strength of game-theoretic design lies in its decentralized approach, which provides a foundational structure for building and managing complex, large-scale networks \cite{goyal2023networks,jackson2008social}. In these decentralized networks, individual agents act based on personal incentives, often with limited or no knowledge of the network’s overall state. This bottom-up approach mirrors real-world systems, where centralized control may be impractical or ineffective.

\begin{figure}
    \centering
    \includegraphics[width=1.0\linewidth]{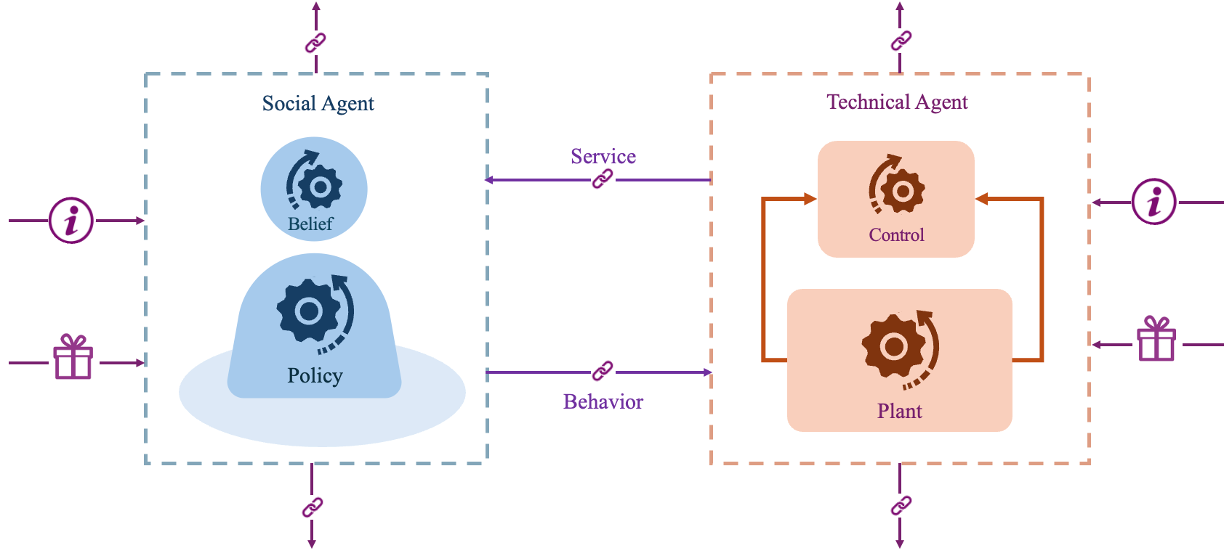}
    \caption{Illustration of Interaction Between a Social Agent in the Human Network and a Machine Agent in the Technical Network: A social agent interacts with the human network and a machine agent within the technical network. Each agent is also connected to other agents within its own network. The machine agent provides specific services to the social agent, while the social agent impacts the machine agent and its network through behaviors such as consumption, usage, or demand patterns. The designer can strategically influence both networks using tools like information design and incentive structures. Information design shapes the structure of information between agents, while incentive design aligns agent actions with broader system goals, creating a coordinated and adaptive socio-technical system.}\label{agents}
\end{figure}

For human agents, game-theoretic design allows for behavior modification to align with system goals. For machine agents, it enables programming diverse agents to follow a coordinated protocol. These agents, whether human or machine, can work collectively to achieve desired outcomes associated with metrics such as efficiency, robustness, resilience, and security. By embedding game-theoretic strategies, designers can anticipate and guide agent interactions, facilitating cooperative behavior even in environments with limited information sharing or direct coordination. Figure \ref{agents} presents an agent-based perspective on socio-technical systems illustrated in Figure \ref{overall}. Within this framework, human agents within human networks engage with machine agents in technical networks, while also interacting with other agents in their respective networks. Each human agent operates as a coupled system, integrating individual belief processes with action processes. Similarly, each machine agent functions as a coupled system, linking control processes with physical processes. Designers can influence various agents through targeted levers across different system components, aiming to optimize system-level performance.

The design of agents is closely linked to control theory, specifically the design of controllers that manage dynamical systems to achieve desirable properties like stability and optimality. While control theory traditionally focuses on governing centralized control systems, game-theoretic agent design introduces a complementary approach that is particularly suited to large-scale socio-technical networks. In these networks, the goal is often to achieve outcomes such as optimal social welfare or collective efficiency, which align with the objectives of control theory. Game-theoretic design operates from the bottom up, creating decentralized agents that make decisions based on local information and personal incentives. This bottom-up approach enables scalability, making it ideal for vast, complex networks typical of socio-technical systems, where centralized control may be impractical. By designing agents to act independently yet cohesively, game-theoretic design facilitates adaptable, resilient, and efficient network behaviors, even in highly dynamic and large-scale environments.

\noindent{\bf Modeling of the Agents} For the socio-technical system illustrated in Figure \ref{overall}, game theory can model diverse interactions within socio-technical networks from the ground up. These interactions can be categorized into several key types. First, interactions occur between agents within the same network, such as those between human agents in human networks or machine agents within technical networks. These intra-network interactions capture the dynamics among similar types of agents and can reveal emergent patterns within isolated sub-systems. Second, interactions take place between agents across different networks. For instance, human agents in the socio-network interact with machine agents in the technical network, bridging the socio-technical divide. These cross-network interactions are crucial for understanding how human and machine agents jointly influence system outcomes.

A third category involves interactions with adversarial agents. Adversarial agents are specifically introduced to evaluate the security, robustness, and resilience of the network. These adversarial entities may be real participants within the network or artificial agents created to assess risk. By engaging human or technical agents with adversaries designed with specific intentions and capabilities, we can measure local security and resilience properties more accurately. Finally, interactions occur between agents and a designer. Here, a designer exerts influence over agents in a controlled way to guide their behavior toward achieving network-wide objectives. This interaction serves as a means of designing and controlling agent actions within the network to align with broader system goals. Each of these interactions takes on distinct forms, and the various games representing them are ultimately composed into a larger framework, referred to as a “meta-game.” This meta-game governs the design and control of the entire socio-technical network, enabling a holistic approach to understanding and managing complex interactions within the system.

\noindent{\bf Control of the Agents} Agents can be controlled in various ways, depending on their nature and function, and these controls can be categorized into three primary paradigms. The first is physical control, which involves managing physical attributes like speed, direction, and other measurable quantities, as seen in robotic agents \cite{zhao2024learning}. The second is cyber control, where the focus is on controlling the information received by agents, such as news broadcasts for human agents or sensor data for autonomous vehicles. The third paradigm is human control, where the objective is to influence perceptions and incentives to guide human behavior in desired directions. Across these paradigms, network structure and information design are fundamental. How agents communicate, physically interact, and gather information from observations and perceptions are critical components of effective system design \cite{li2022role}.

A key connection between control theory and game-theoretic design emerges through the use of dynamic game frameworks to model and guide agent behavior in evolving environments \cite{huang2020dynamic,bacsar1998dynamic}. In dynamic games \cite{BZ18}, agents interact over time within changing environments and face uncertainties. Agent behaviors are characterized by adaptive feedback loops, where decisions continuously adjust based on environmental conditions. Information flow becomes particularly crucial in these scenarios, as agents make real-time decisions with limited or noisy information about others’ actions. The flow and structure of information directly shape agents' strategic choices, influencing the overall system’s resilience and robustness. Bridging control and game-theoretic design achieves a unified approach to achieving individual dynamic agents and ensuring the stability and efficiency of the entire system. On the individual level, agents must operate effectively within their local environments, maintaining stability in response to changing conditions and achieving their own performance goals. At the system level, however, the design must prioritize overall stability, resilience, and system-wide metrics \cite{zhu2015game,zhu2024disentangling}.

\subsection{The Underlying Philosophy of Agent-Based Game-Theoretic Design in Socio-Technical Networks}

The agent-based game-theoretic design of socio-technical networks embodies a dual philosophy \cite{ponte2016holism,verschuuren2017holism}: reductionistic design and holistic control. On one side, game-theoretic design takes a reductionist approach, where the whole system is decomposed into modular components or agents. By breaking down complex, large-scale networks into manageable agents, this approach allows designers to handle intricate interdependencies and diverse functions within the system. On the other side, the design aims to achieve high-level system objectives—such as efficiency, security, and resilience—which are often prescribed at a system-wide level. The reductionist design of individual agents must, therefore, be aligned with these holistic goals, ensuring coherence between component-level actions and overall system performance. Establishing such coherence is fundamental to the principles guiding game-theoretic design in socio-technical networks.

Achieving coherence between agent-level design and system-level objectives requires a framework to bridge them. Designers need to assess how individual agent behaviors impact system-wide metrics, making it essential to monitor the alignment of component actions with system goals. Game-theoretic analysis provides this bridge by offering a structured framework to predict system-level behaviors through equilibrium concepts. The equilibrium, depending on the application and structure of the network, enables designers to forecast the outcomes of individual actions within the larger system. Various solution concepts within game theory offer tools to assess and develop performance metrics. For instance, in a security context, equilibrium analysis between a defender agent and an attacker can yield risk metrics, while in robustness analysis, saddle-point equilibria between the system and external disturbances inform robustness metrics.

Game theory serves as a bridge that enables reductionist designers to account for the holistic impact of individual agent designs on the system's objectives. Meanwhile, holistic system designers must shape the architecture, including hierarchies, network structures, and resource allocations, to ensure that agent-level designs contribute to the system’s high-level goals. Frameworks such as Stackelberg games, equilibrium-constrained optimization, and mechanism design theory play a central role in achieving this alignment. Holistic designers must understand how agents respond to these structures at equilibrium and ensure that top-down control strategies foster the intended system-wide behavior.

Ensuring coherence becomes more challenging under conditions of uncertainty, adaptive requirements, and emergent properties such as resilience and security. These complex requirements demand clear, quantifiable metrics to guide system and agent design. Despite these challenges, game theory—with its rich array of tools and methodologies—provides a means to develop advanced techniques that foster coherence in dynamic, complex systems. For instance, game-theoretic tools can incorporate learning and adaptation, enabling agent designs to evolve in response to an uncertain environment while staying aligned with system goals.

\begin{figure}
    \centering
    \includegraphics[width=0.8\linewidth]{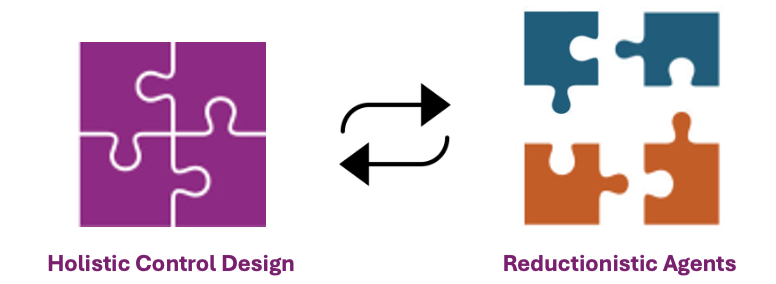}
    \caption{The holistic control design must align consistently with the reductionist behaviors of individual agents. Game theory, inherently a reductionist approach, focuses on designing and analyzing individual agent behaviors, while control theory provides a holistic framework to achieve overarching system goals. Game-theoretic control offers a cohesive approach that bridges these two perspectives, integrating the detailed evaluation and synthesis tools of reductionist models with the coordination and control mechanisms of holistic design. This combined framework ensures that individual agent actions are aligned with the broader system objectives, creating a unified and adaptive socio-technical system.}\label{coherence}
\end{figure}

This coherence between reductionism and holistic control, illustrated in Figure \ref{coherence}, is where game theory and control theory intersect, together forming the foundation for a new system design paradigm. Designing socio-technical networks requires this paradigm shift and the convergence between control and game theory to address the unique demands of these complex, interconnected systems. For example, in a smart grid, it is insufficient to simply control each subsystem, such as energy generation or distribution, in isolation. The system must account for the interplay between independent agents (e.g., consumers, generators, and grid operators) who each respond to incentives, environmental conditions, and their own objectives. By integrating game-theoretic strategies, designers can predict how these agents will behave collectively, while control theory enables the coordination of these actions to maintain grid stability, efficiency, and resilience.

\subsection{Organization of the Letter}
 
 This letter provides an overview of game-theoretic design approaches. In Section II, we explore foundational frameworks used in agent design, focusing on Stackelberg-type game frameworks and mechanism design theory, which has been widely applied to settings like auctions and market structures. Section III examines the challenges of designing socio-technical systems, addressing issues such as human behavioral dynamics, uncertainty quantification, and scalability. In Section IV, we present emerging paradigms in game-theoretic design, including mean-field design, learning-based design, population-based design, and adversarial design. These approaches are applied to critical areas such as misinformation management in social networks, resilience in industrial control systems, and congestion control in infrastructure networks. We close the letter with the concluding remarks of Section V.

\section{Baseline Approach}
\label{base}
This section introduces the foundational frameworks for game-theoretic design, focusing on two primary models. The first is the Stackelberg framework, which employs a hierarchical bilevel optimization structure. The second is the mechanism design framework, which also features a bilevel structure and can be viewed as a specific form of Stackelberg game, often applied in market design to establish rules, including allocation and payment policies. Both frameworks have broad applications across engineering and economics and have been extended to multi-player, dynamic environments, where  insights from control and system theories become invaluable. Emerging challenges are discussed in this section.

\subsection{Stackelberg Games}

Stackelberg games provide a foundational framework for designing agents in socio-technical networks, particularly when there is a hierarchical or leader-follower structure in decision-making. Originating from economics \cite{von2010market}, Stackelberg games model scenarios where a leader (or central authority) makes a decision first, anticipating the responses of other agents, known as followers. In socio-technical systems, this approach enables the design of mechanisms that coordinate agents in the network or agents that can strategically influence or respond to the actions of other agents within the network, effectively managing system dynamics and optimizing outcomes.

To systematically formalize the Stackelberg game framework, we begin by introducing basic notations and concepts that establish the hierarchical interaction between agents, typically referred to as the leader and the follower. Consider a scenario where a leader, often a central authority or a key stakeholder, makes a decision or selects an action that influences other agents within a network. Let \( x \in X \) denote the action or decision taken by the leader, where \( X \) represents the leader’s feasible action space.  The action In this hierarchical structure, the leader's choice of \( x \) precedes and directly impacts the response of the follower agent in the system.

Upon observing the leader’s decision, the follower chooses its own action in response. Let \( y \in Y(x) \) represent the action chosen by the follower, where \( Y(x) \) denotes the set of feasible actions available to the follower, conditioned on the leader’s choice \( x \). Thus, the follower's available choices depend on and are constrained by the leader's prior action, creating a structure where the follower adapts based on the leader’s influence. Note that the action spaces of the players can take different mathematical forms based on the context and the applications. Each agent, both the leader and the follower, seeks to maximize an individual payoff, represented by specific payoff functions that depend on both agents' actions. The leader’s payoff function is denoted as \( U_L(x, y) \), which captures the outcome or benefit to the leader based on its choice \( x \) and the follower's response \( y \). Similarly, the follower's payoff function is denoted as \( U_F(x, y) \), reflecting the outcome or benefit to the follower that depends on both the leader’s decision \( x \) and the follower's action \( y \).

Given the leader's decision \( x \), the follower aims to choose the action \( y \) that maximizes its own payoff \( U_F \). Formally, this choice is represented by the best response function \( y^*(x) \), which is assumed here to be unique for each $x\in X$, defined as:
\[
y^*(x) = \arg \max_{y \in Y(x)} U_F(x, y)
\]
This formulation implies that, for a given leader action \( x \), the follower will select the action \( y^*(x) \) that provides the highest possible payoff, given the constraints imposed by the leader’s decision. In this way, the follower's behavior can be anticipated based on its objective to maximize its own payoff.

The leader, in turn, anticipates this unique best-response behavior from the follower when selecting its action. Knowing the follower’s best-response function \( y^*(x) \), the leader chooses its own optimal action \( x^* \) by solving the following optimization problem:
\[
x^* = \arg \max_{x \in X} U_L(x, y^*(x))
\]
Here, \( x^* \) represents the leader’s optimal decision, taking into account the follower's anticipated response \( y^*(x) \). This process captures the leader's strategic anticipation of the follower's behavior and allows the leader to optimize its payoff based on expected follower reactions.

\begin{figure}
    \centering
    \includegraphics[width=0.8\linewidth]{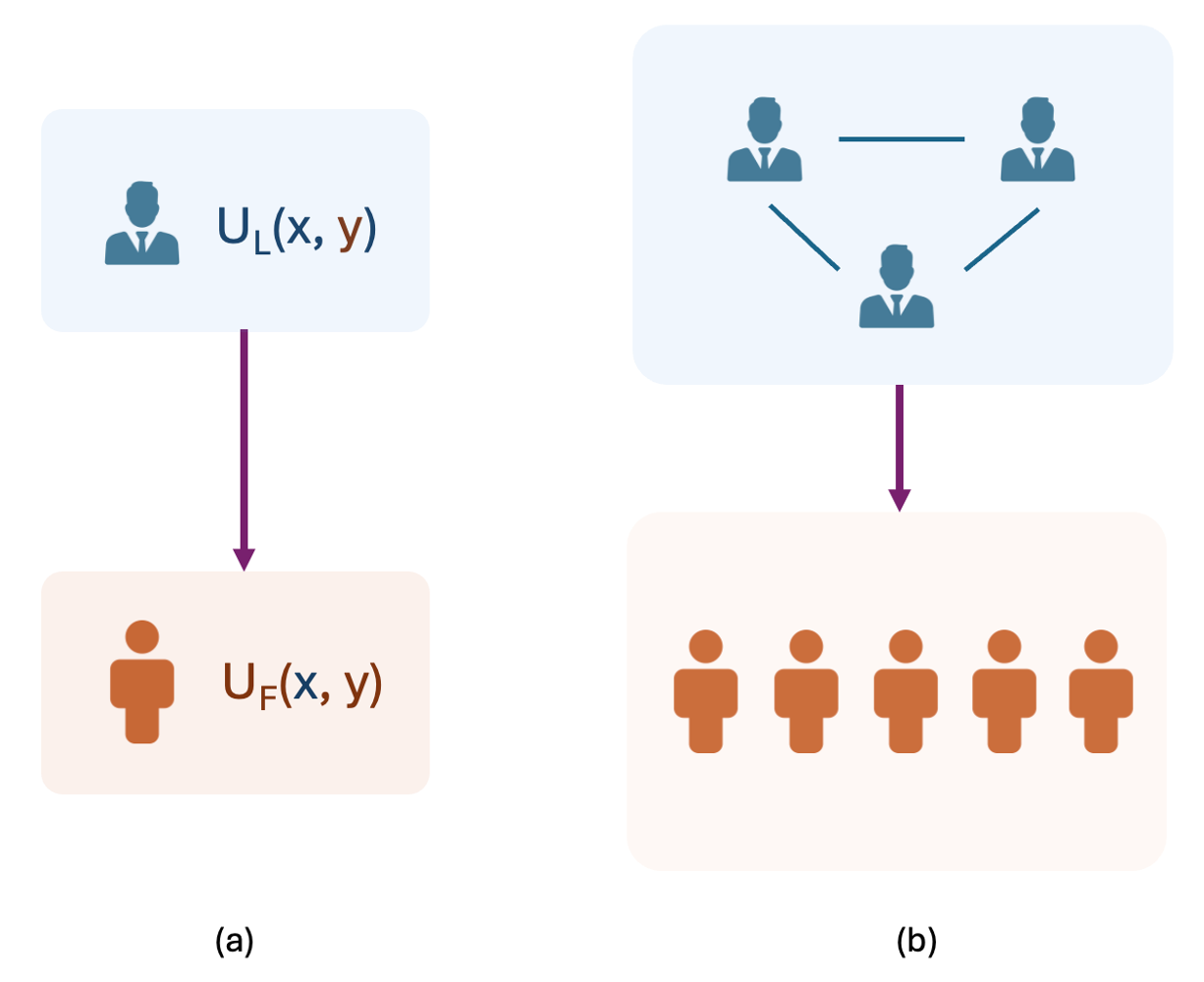}
    \caption{Illustration of the Stackelberg Game Framework: (a) In a basic Stackelberg game framework, there are two agents: a leader and a follower. The leader makes the first move, setting the stage, and the follower responds based on the leader's action, optimizing their own outcome within the constraints set by the leader’s decision. (b) This framework can be extended to include multiple leaders and multiple followers, creating a more complex system with interactions both within each group and between groups. Leaders coordinate their actions, considering potential responses from followers, while followers adjust based on both leader actions and interactions with other followers. This multi-agent Stackelberg model captures the layered decision-making and interdependencies in complex systems.}\label{Stackelberg}
\end{figure}

The combination \((x^*, y^*(x^*))\) constitutes a \textit{Stackelberg equilibrium}, where both the leader's optimal action \( x^* \) and the follower's corresponding best response \( y^*(x^*) \) satisfy each agent's strategic objectives. This equilibrium formalizes a stable outcome in which the leader maximizes its payoff while taking into account the follower's best response, and the follower optimally responds to the leader's chosen action. This baseline framework is summarized in Figure \ref{Stackelberg}(a). More details of Stackelberg equilibrium can be found in \cite{bacsar1998dynamic},  including extension of this formulation to the case when the follower's best response is not unique.. The baseline two-player game model can be extended to a multi-leader, multi-follower framework, which is illustrated in Figure \ref{Stackelberg}(b). In this setting, interactions occur not only  between leaders and followers but also among leaders and among followers. Such dynamics are typical in socio-technical systems (see \cite{chen2019game,albaba2019modeling}), where individuals interact within a network and are simultaneously influenced by various policymakers or influencers in distinct ways. The Stackelberg equilibrium offers a mechanism for leaders to influence and stabilize system dynamics by strategically guiding follower behaviors, aligning local actions with broader system objectives.

 The traditional Stackelberg game framework can be extended to \textit{dynamic Stackelberg games} \cite{bacsar1986tutorial}, which are characterized by leader-follower interactions that unfold over multiple stages or continuously over time. In a dynamic setting, both leader and follower agents make sequential or ongoing decisions, adapting their strategies based on observed or anticipated changes in the system and in each other’s actions. This dynamic approach provides enhanced flexibility for modeling complex, multi-round interactions, which is particularly useful in real-world systems where decision-makers continuously adjust to new information and changing conditions. A key feature of dynamic Stackelberg games is \textit{sequential decision-making over time}. Unlike static Stackelberg games, where the leader makes a one-time decision, dynamic Stackelberg games involve repeated or continuous decisions from both the leader and followers. The leader’s actions may evolve over time, allowing adjustments in response to the follower’s behavior or environmental factors. This sequential structure makes dynamic Stackelberg games ideal for modeling real-time systems, such as financial markets, adaptive security protocols, and smart infrastructure management. Dynamic Stackelberg games also involve \textit{time-dependent payoffs and strategies}. In dynamic settings, payoffs for both the leader and followers can vary over time, influenced by factors like environmental shifts, market trends, or changes in agent states. As payoffs change, both the leader and follower adjust their strategies to reflect these new values, often using dynamic optimization techniques, such as differential game theory or optimal control, to account for the time-dependency of strategies. Another key component is the use of \textit{feedback mechanisms}. Dynamic Stackelberg games frequently employ \textit{feedback strategies}, where the leader observes the follower’s actions in real time and adjusts decisions accordingly. This contrasts with \textit{open-loop strategies}, in which the leader commits to a fixed path without considering the follower’s ongoing behavior. Feedback mechanisms enhance adaptability, allowing the leader to respond to unexpected changes and making the system more resilient to volatility. In addition, \textit{state-dependent strategies and control} are essential in dynamic Stackelberg games. Here, the system's state (e.g., network load, asset prices, or security threat levels) significantly influences the agents' decisions. Strategies for both the leader and follower are conditioned on the evolving state, enabling agents to adapt their actions based on current conditions. Interested readers can refer to \cite{basear1984feedback,bacsar1998dynamic} for more details. This feature makes dynamic Stackelberg games particularly relevant to fields like cyber-physical systems \cite{huang2020dynamic,zhao2022stackelberg,zhao2023stackelberg}, where decisions depend on factors such as network congestion or resource availability.



The Stackelberg game framework  provides a structured way to model and optimize hierarchical interactions, making it especially useful for designing decentralized and adaptive agent-based systems. By formalizing how agents with differing incentives and information interact, Stackelberg games allow system designers to influence complex socio-technical networks and improve overall network performance by setting effective policies and incentives.
Dynamic Stackelberg games are applicable in various fields. In \textit{smart grid management}, for example, a utility provider (leader) may dynamically set pricing strategies in response to real-time fluctuations in energy demand and supply, and consumers (followers) adjust their consumption accordingly (see \cite{maharjan2013dependable,chen2017stackelberg,GBD-G16,ALCB21}). Similarly, in \textit{traffic and congestion control}, traffic authorities (leaders) set adaptive toll prices or traffic signals based on current traffic patterns, prompting drivers (followers) to adjust their routes and travel schedules (see \cite{yang2007stackelberg,korilis1997achieving}). In \textit{cybersecurity and defense systems}, defenders (leaders) dynamically adjust security protocols in response to real-time threat assessments, while attackers (followers) adapt their strategies to the defenders' moves (see \cite{pan2022poisoned,pawlick2016stackelberg}).

\subsection{Mechanism Design Frameworks}

Mechanism design theory is another foundational approach in game theory focused on crafting rules and incentives that guide independent, self-interested agents—each with private information and individual preferences—toward desired outcomes. This approach considers two core elements: the \emph{environment} and the \emph{mechanism}. The environment, beyond the designer’s control, includes the participants, possible outcomes, and participants' \emph{types}—which encapsulate their preferences, information, and beliefs. The \emph{mechanism} is the structured set of rules that defines the choices available to participants and maps these choices, or ``messages,'' to specific outcomes based on participants' revealed preferences.

In mechanism design, the central goal is to achieve \emph{incentive compatibility}, meaning agents are motivated to disclose truthful information that aligns with their interests. This compatibility enables the mechanism to reach optimal outcomes while aligning participants’ actions with the designer's objectives. A powerful tool in mechanism design is the \emph{Revelation Principle}, which simplifies complex design problems by focusing on \emph{direct mechanisms}, where participants report their types or preferences directly, facilitating transparent and efficient analysis. Through these principles, mechanism design theory supports the development of robust frameworks for achieving fair, efficient outcomes in environments where information is decentralized and participants act independently. Interested readers can refer to \cite{fudenberg1991,maschler2013game} for details.

In socio-technical networks, mechanism design frameworks have wide applications. Auction-based resource allocation mechanisms, for example, are used to manage scarce resources like bandwidth or energy in socio-technical networks. Here, auctions ensure that resources are allocated efficiently among agents with competing demands. The Vickrey-Clarke-Groves (VCG) mechanism is a common approach, incentivizing agents to reveal their true values, maximizing social welfare and fairness in sectors such as telecommunications \cite{chen2023qos,farooq2018optimal,zhu2008state}, cloud computing \cite{chen2016optimal}, and decentralized energy systems \cite{chen2022transactive}. Market-based structures are also common in distributed systems, where agents act as buyers and sellers. Price-setting mechanisms in these markets help balance supply and demand, aligning individual actions with system-wide efficiency, particularly in decentralized energy grids \cite{chen2016game,zhu2012game}, ride-sharing services \cite{ma2022spatio,hong2022optimal}, and supply chain networks \cite{nagurney2010sustainable,leng2005game}.

In environments where agents need to be paired—such as job markets or school placements—matching mechanisms provide stable, mutually beneficial assignments. The Gale-Shapley algorithm \cite{gale1962college}, a widely-used matching method, ensures stable matches that prevent incentives for misrepresentation, supporting fairness and stability across the system. Mechanism design also addresses collective decision-making through voting mechanisms \cite{jackson2002coalition}, which aggregate agents' preferences to produce fair, representative outcomes. Voting frameworks are critical in policy selection, group resource allocation, and distributed decision-making in social networks, enhancing social welfare and reducing the risk of manipulation.

Applications of mechanism design in socio-technical networks include traffic and congestion management \cite{lombardi2021model}, where mechanisms such as dynamic toll pricing incentivize drivers to avoid peak times or take alternative routes, reducing delays and improving efficiency. In a similar vein, mechanism design has been applied to communication networks for fair bandwidth allocation and congestion control through also differential pricing \cite{SB07,SB11}. In social networks and online platforms \cite{yang2023designing,zhang2019optimal}, mechanism design can create incentives that encourage accurate information-sharing and discourage misinformation, establishing rewards for verified content and penalties for spreading false information. This approach is also instrumental in decentralized energy systems \cite{zhu2012game,zhu2012differential}, where mechanism design can support grid stability by encouraging small-scale producers and consumers to align their energy use with system demands.

Recent efforts have also investigated dynamic mechanism design theory (see \cite{zhang2023dynamic,chen2021dynamic}) addressing the complexities of time-evolving environments, where agents' information, preferences, or actions change dynamically. This framework is essential for managing socio-technical systems where agents interact persistently and where adaptability is key to achieving desired system outcomes, such as efficiency, robustness, or resilience. In dynamic mechanism design, the principal, often a system regulator or network administrator, must create rules and incentives that not only guide agents’ actions in the present but also anticipate and adapt to future changes in agent behavior and information. This approach is particularly useful in settings where agents' private information, such as preferences, costs, or objectives, can evolve over time, potentially influencing the agents’ incentives to participate in or cooperate with the system’s rules. By designing mechanisms that account for these changes, the principal ensures that agents remain motivated to act in ways that support collective objectives, even as their private states shift.

Key to dynamic mechanism design is the concept of dynamic incentive compatibility \cite{zhang2022incentive}. Unlike in static settings, where incentives are often one-off, dynamic settings require that agents have incentives to behave cooperatively at every stage. This involves designing mechanisms that are truthful over time, meaning agents are motivated to continuously reveal private information accurately and take actions aligned with the system’s objectives. For instance, in a smart grid, energy consumers might adjust their usage based on price signals that are updated in real-time, incentivizing continuous alignment with grid efficiency goals.

To achieve dynamic incentive compatibility, dynamic mechanism design often employs tools like state-dependent contracts or future-oriented rewards. These tools ensure that agents consider the long-term consequences of their actions, as their future payoffs may depend on maintaining truthful and cooperative behavior over time. For example, in a decentralized transportation network, a mechanism might reward drivers for consistent off-peak travel behavior, not just a one-time response, thereby reducing congestion sustainably.

\subsection{Design Challenges in Socio-Technical Networks}
Game-theoretic design in socio-technical networks faces significant challenges due to the complexities of large-scale interactions, human behavior dynamics, and evolving system states. Addressing these issues requires extending foundational frameworks to effectively tackle the following challenges.

\subsubsection{Handling Paradoxes}
In socio-technical systems, certain paradoxes can arise when local optimizations unintentionally lead to suboptimal or even adverse outcomes at the system level. One prominent example is Braess' paradox \cite{nagurney2021braess}, where adding resources (e.g., a new road in a transportation network) can increase overall congestion rather than alleviate it. Dynamic mechanism design addresses these paradoxes through careful game-theoretic analysis, structuring incentives that prevent agents from taking actions that inadvertently harm collective outcomes (see \cite{yang2024prada,yang2023strategic,altman2006survey}). By designing mechanisms that anticipate and mitigate these paradoxes, it is possible to guide agents' behaviors in ways that optimize the system holistically, even when local choices may appear beneficial in isolation in traffic and communication networks \cite{altman2001avoiding,altman2006survey}. 

\subsubsection{Human Bounded Rationality}
In real-world settings, human agents often make decisions within the limitations of bounded rationality—they may not process all available information optimally or may rely on heuristics rather than rational calculation. To accommodate this, dynamic mechanism design can incorporate bounded-rational frameworks that account for human cognitive limits \cite{simon1990bounded,huang2023review}. For example, rational inattention theory aims to model situations where agents selectively process only certain information due to cognitive or resource constraints \cite{sims2003implications}. Prospect theory, on the other hand, addresses cognitive biases by capturing how agents evaluate potential losses more heavily than equivalent gains, thereby modeling risk-related decision-making deviations from purely rational behavior \cite{kahneman2013prospect,SSB20}. Additionally, risk aversion can be analyzed through frameworks of risk-sensitive decision-making, where diverse risk and utility measures are employed to capture the varying risk preferences of decision-makers \cite{TB21}. This approach allows for a nuanced understanding of behavior, accounting for differences in how individuals weigh potential losses and gains.  Incorporating these frameworks ensures that mechanisms remain effective even when agents operate under limited information-processing capabilities, enhancing the system's robustness to human cognitive constraints. Furthermore, it is also possible to design unconventional mechanisms that leverage specific features of bounded rationality to achieve or encourage desirable behaviors. In this view, bounded rationality serves as an additional design lever, guiding agents toward outcomes that align with the designer’s objectives \cite{huang2023cognitive,huang2024psyborg+,huang2022radams}.  

\subsubsection{Scalability}
Scaling dynamic mechanism design to large socio-technical networks presents significant challenges due to the computational complexity and the large number of interacting agents. Approaches like mean-field games \cite{basar2020recent,caines2021mean}, multi-agent learning \cite{li2022role,li2022confluence,ZYB21}, and federated learning \cite{kairouz2021advances,zhang2021survey} are useful techniques to handle these issues. Mean-field games, for instance, approximate the behavior of large populations by modeling the average effects of agent interactions rather than tracking each agent individually. Multi-agent learning enables decentralized agents to adaptively learn optimal strategies over time, while computational optimization algorithms efficiently process large datasets and complex models. These techniques together allow the mechanism to remain computationally feasible and effective even as network size and complexity increase.

\subsubsection{Uncertainty Quantification and Management}
In dynamic socio-technical environments, agents and system designers often operate under incomplete or uncertain information, such as fluctuating demand, changing preferences, or unpredictable external influences. To manage this uncertainty, dynamic mechanism design leverages learning-based methods that allow mechanisms to adapt based on observed data. Machine learning models, for example, can predict agent behavior patterns or environmental shifts \cite{li2023self,li2024multi}, informing real-time adjustments to incentives. By incorporating methods that continuously learn from data, dynamic mechanism design can respond to uncertainty with adaptive mechanisms, maintaining robust performance even when facing incomplete information.

\subsubsection{Threat Models}
In many applications, socio-technical networks are vulnerable to adversarial attacks or disruptions, such as cyber-attacks, fraud, or manipulation by malicious agents. Designing robust mechanisms that can withstand these threats is a core challenge in dynamic mechanism design. By incorporating threat models, designers can anticipate potential adversarial behaviors and create mechanisms that are resilient to disruptions. Techniques like resilient control \cite{zhu2015game,ishii2022security}, robust optimization \cite{ben-tal2009robust,bacsar2008h}, and game-theoretic risk management \cite{zhu2024foundations,rass2018game,chen2019interdependent}  play a critical role in this process, ensuring that mechanisms can detect, mitigate, and adapt to malicious actions, enabling the system to maintain functionality and stability, even in the presence of adversarial forces.

\section{New Paradigms in Game-Theoretic Design}
\label{new}
 The challenges in game-theoretic design are diverse and complex, stemming from the interplay of human factors, large-scale interactions, information uncertainties, and potential adversarial behaviors. 
 This section explores several innovative design paradigms to address these challenges. These paradigms can be integrated within frameworks such as Stackelberg games or mechanism design by strategically embedding design variables. Moving beyond traditional incentive or payment rule design, these new approaches extend into domains of information structuring, network configuration, and learning-based designs. These expanded frameworks target critical system properties, such as security, robustness, and resilience, that go beyond conventional metrics of efficiency or social welfare.

\subsection{Information Design}

Information Design is a crucial approach within game theory and mechanism design, focusing on structuring the information available to agents within a system to influence their decisions and behaviors. In socio-technical networks, where agents make decisions based on the information they receive, information design plays a powerful role \cite{bergemann2019information,li2022role}. By strategically crafting the flow and structure of information, designers can guide agents toward actions that align with collective goals such as efficiency, fairness, stability, and security. Information design differs from direct incentive manipulation \cite{TB24}. Instead, it subtly alters agents' perceptions and beliefs, which in turn influences their actions without changing their underlying payoffs. Information design centers around constructing information structures and shaping agent beliefs. Designers control the type, timing, and quality of information available to agents, which influences their beliefs about the system state and the actions of others \cite{li2024decision}. In socio-technical networks, agents may have incomplete knowledge of the system, potential outcomes of their actions, or even the intentions of other agents. By structuring what agents know and when they know it, designers can guide agents toward desirable behaviors. For example, selectively disclosing traffic information in a transportation network can influence drivers to avoid congested routes and thus overall traffic flow without imposing mandatory controls.

One of the primary methods of implementing information design is through signals, structured pieces of information sent to agents to influence decisions without dictating specific actions. For instance, in a power grid, providing consumers with signals about peak usage times can encourage energy-saving behaviors during high-demand periods, thereby reducing grid strain. Signals may be deterministic, offering clear and direct information, or stochastic, offering probabilistic guidance. The choice of deterministic versus stochastic signals depends on the desired effect and the network's goals. In many socio-technical networks, full transparency can be neither feasible nor beneficial \cite{li2023price}. Revealing too much information can lead to undesirable behaviors. Selectively revealing partial information can influence behavior in ways that support system goals.  For instance, in financial networks, withholding certain market data can prevent herding behaviors that could destabilize markets. By controlling what agents know and when they know it, designers can influence decision-making while reducing uncertainty and preventing potentially harmful actions. 


Applications of information design are widespread in socio-technical networks. In transportation networks \cite{massicot2019public}, information design is instrumental in managing congestion. Providing drivers with real-time traffic updates or estimated travel times allows city planners to influence route choices, distributing traffic more evenly across available roads. In some cases, withholding information can prevent overreaction; for instance, if all drivers are informed of the optimal route simultaneously, new congestion points may emerge. By designing information structures that guide but do not dictate decisions, planners can improve traffic flow and reduce delays. In cybersecurity, information design guides user behaviors by providing carefully constructed alerts about potential threats, encouraging more secure practices such as using strong passwords or avoiding risky websites. Misinformation, or "honey tokens," can also divert potential attackers away from valuable assets, providing deceptive signals to protect critical resources without revealing genuine vulnerabilities.

In energy networks, information design shapes consumer behavior through well-timed signals about energy usage, pricing, and renewable energy availability. By providing consumers with real-time or forecasted data on energy prices and demand peaks, designers encourage shifts in energy consumption patterns that promote grid stability \cite{aid2023signalling}. Signals can prompt users to consume less energy during peak periods or to use renewable energy sources, thereby enhancing efficiency and resilience across the grid. Similarly, during health crises or emergencies, information design is essential for managing public behavior. For example, in a pandemic, timely updates on case numbers, safety protocols, or vaccine availability can encourage individuals to follow public health guidelines \cite{liu2022eproach}. By releasing information incrementally or with specific framing, public health agencies can encourage cautious and cooperative behaviors that contribute to community health and safety.

Implementing information design in socio-technical networks involves several challenges. One key challenge is balancing transparency and control. Too much transparency can lead to paradoxical outcomes \cite{zhu2010price,li2023price}, while too little can lead to distrust or inefficiency. Effective information design requires understanding how different levels of information disclosure impact agent behavior and balancing transparency to achieve the desired influence without causing unintended consequences. Another challenge is adapting to dynamic environments \cite{zhang2023stochastic,liu2023information,velicheti2023strategic}, as socio-technical networks often evolve over time. Information structures that are effective under one set of conditions may have unintended effects under another, requiring continuous monitoring and adjustment to keep information design aligned with network objectives. Additionally, agents may respond unpredictably to new information structures, particularly if they attempt to ``game" the system. Designers must anticipate these strategic reactions and discourage gaming behavior, while still guiding agents toward beneficial actions \cite{VBEB24}.

Finally, information design can raise ethical concerns, especially when it involves withholding information or presenting partial truths. In sensitive domains such as healthcare or social media, designers must carefully weigh the ethical implications of their choices, balancing effectiveness with respect for agent autonomy and privacy. Establishing transparent guidelines on when and why information is withheld is essential for maintaining trust and adhering to ethical standards. In \cite{zhu2023doctrine}, the doctrine of cyber effect introduces five ethical principles—goodwill, deontology, no-harm, transparency, and fairness--to guide the ethical design of cyber systems, including recommendation engines, cyber deception mechanisms, and information design frameworks. These principles serve as a foundation for creating systems that are both effective and ethically responsible.




\subsection{Network Design}

Network design represents a powerful paradigm within game-theoretic control, enabling the management of agent interactions by shaping the structure of the network. In socio-technical networks, multiple interconnected layers typically exist: a human layer involving social and behavioral interactions, and a technical layer that provides the infrastructure supporting these interactions. Designing connectivity within the human layer can be achieved by establishing communication and signaling channels that guide behavior and facilitate information flow. Meanwhile, connectivity in socio-technical interactions can be orchestrated through service delivery; humans interact with the infrastructure to access specific services, while the infrastructure itself can adjust elements like pricing and quality of service to attract or deter human engagement. The infrastructure layer’s network topology—how components like data centers, transportation routes, and communication links are connected—also plays a crucial role in shaping agent behaviors.

Transportation systems provide a clear example of this multi-layered network design. Users may be encouraged to use different transportation modes at different times based on real-time conditions and personal preferences. The design of interconnected transportation networks, such as multi-modal hubs, enables users to switch between different routes or modes to reach their destinations, thus distributing demand across the network. Game-theoretic pricing mechanisms, such as dynamic tolls or off-peak discounts \cite{lombardi2021model}, can influence user choices to manage congestion more effectively. The interdependency between human networks and the transportation network itself means that any changes in infrastructure—like altering connectivity or adjusting service quality—will influence human movement patterns and vice versa. 

Game-theoretic design can be applied at individual layers or across these interconnected layers \cite{chen2019game}. In a single-layer approach, designers might focus on influencing only human behavior (e.g., encouraging route adjustments based on real-time information). In a cross-layer approach, however, game-theoretic strategies account for both human and infrastructure layers simultaneously, designing incentives and network configurations that influence how agents interact within and between layers. For example, infrastructure adjustments, like expanding transit options during peak hours, can be combined with user incentives, such as discounted fares, to collectively reduce congestion across the network.

This cross-layer approach is particularly effective for resilience-focused socio-technical networks. In emergency response networks, for instance, designing robust connectivity across both human and infrastructure layers can ensure effective information flow, even in disrupted conditions. Agents, whether human responders or automated systems, can be incentivized to make strategic decisions that stabilize the network under stress. By utilizing game-theoretic network design, socio-technical networks achieve a holistic balance of security, robustness, and user satisfaction, aligning agent behaviors with system-wide objectives while fostering adaptability across these interconnected layers. In \cite{zhu2015game,chen2019control}, several advanced tools for game-theoretic design of socio-technical networks are introduced, particularly for enhancing resilience, security, and adaptability within multi-layered, interdependent systems. The Games-in-Games framework, which leverages multi-layered Stackelberg and Nash games, is designed to enhance security in adversarial environments and optimize connectivity within interdependent networks. Building on this, Meta-Equilibrium Solutions coordinate cross-layer objectives, enabling cohesive defense strategies that align individual and system-wide goals within these complex socio-technical networks.

\subsection{Population Design}
Population design framework in socio-technical networks leverages principles from population games \cite{sandholm2010population} to manage and influence large groups of agents interacting within a shared environment. Rather than focusing on individual agent decisions, population design considers the behavior and dynamics of the entire population, allowing for scalable interventions that impact the system as a whole. By employing population-wide strategies such as mean-field game theory \cite{basar2020recent} and evolutionary game theory \cite{sandholm2020evolutionary}, designers can influence collective behaviors, shaping the outcomes of complex socio-technical systems effectively. 

In population games, agents repeatedly make decisions influenced by the strategies of others within the population. Unlike traditional game theory, which typically models interactions between a fixed number of agents, population games scale this concept to large groups, making them well-suited for socio-technical networks where the number of interacting agents can be vast (e.g., users in a social network, vehicles in a transportation network, or devices in the Internet of Things). Each agent's payoff depends on their own strategy as well as the aggregate strategies of the entire population, creating a feedback loop where individual choices shape the collective outcome, and vice versa.

The core objective of population design in socio-technical networks is to guide the population's collective behavior towards desirable outcomes, such as reduced congestion, improved security, or enhanced resilience. This approach is beneficial for managing large-scale networks, as it enables the system designer to apply population-wide strategies rather than controlling or incentivizing each individual agent separately. Mean-field game theory is a powerful tool within the population design framework, allowing designers to approximate the behavior of large populations by modeling the average effect of interactions among agents. In mean-field games, each agent’s decision is based on the ``mean field," an aggregated representation of the actions and states of the entire population. This mean-field approximation enables system designers to predict and influence collective behavior by implementing policies that alter the mean field.

Using mean-field game theory, leadership-driven strategies can be designed to guide population-wide behavior in socio-technical networks \cite{SBB24,farooq2017secure,farooq2018secure}. For example, in a transportation network, city planners can adjust incentives or provide information (e.g., on congestion patterns or toll prices) that influence the average driving behavior, reducing peak-time congestion without needing to monitor or control individual drivers \cite{ameli2022departure,huang2021dynamic}. Similarly, in smart grids, mean-field models can be used to predict and influence energy consumption by setting dynamic pricing schemes that reflect population-wide consumption trends, encouraging users to shift their energy use during peak times \cite{zhu2013multi,couillet2012electrical}.  The scalability of mean-field game theory makes it ideal for large socio-technical systems, as the designer does not need to model each agent individually. Instead, by shaping the mean field through strategic incentives or informational signals, designers can nudge the population toward behaviors that support system-wide goals.

\subsection{Learning-based Mechanism}
Learning-based mechanism design is an advanced approach that combines design principles with adaptive learning techniques to create dynamic, responsive incentive structures \cite{dai2024incentive,scheid2024incentivized}. In socio-technical networks, where both the environment and agents' behaviors are often unknown and change over time, learning-based mechanism design enables adaptive systems that can respond intelligently to evolving conditions and maintain alignment with system-wide objectives. This approach is particularly valuable in complex networks such as urban transportation, smart grids, and information networks, where fixed incentives or static designs may be insufficient for achieving sustainable outcomes.
 
At the core of learning-based mechanism design is incentive mechanism design, where incentives are structured to motivate agents to behave in ways that benefit the overall system. Unlike static incentive mechanisms, learning-based mechanisms are adaptive, meaning that they continuously update incentives based on observed agent behaviors and environmental conditions. This adaptability is crucial in environments that are highly dynamic or uncertain, where pre-defined incentives may not effectively guide agent behavior as conditions change. In a transportation network, for example, real-time adjustments in toll pricing can influence driver behavior more effectively than a fixed toll rate, especially as traffic patterns fluctuate due to weather, road conditions, or changes in demand.


Learning-based mechanism design relies on creating frameworks that enable the system to learn and optimize incentive structures over time. This is typically done through machine learning models, such as reinforcement learning, which allow the mechanism to explore different incentive strategies and observe their effects on agent behavior. Over time, the system learns which incentives are most effective at achieving network-wide goals, such as reducing congestion, conserving energy, or improving security. In reinforcement learning, for example, the system (or ``agent" in the learning framework) adjusts incentives based on feedback from the environment. The system receives "rewards" based on how well the agents' behaviors align with desired outcomes, and it uses this information to optimize future incentive decisions. This iterative learning process enables the mechanism to adapt to complex and uncertain environments without requiring detailed knowledge of agent preferences or behaviors in advance. In a transportation context, such a learning-based mechanism might adjust tolls in real time based on traffic flow data, maximizing the reward of reduced congestion and improved travel times.

Reward shaping is an essential tool in learning-based mechanism design \cite{laud2004theory}, where rewards are structured to gradually guide agents toward desired behaviors. Reward shaping involves modifying the reward function in a way that reinforces incremental steps toward the target behavior, rather than only rewarding the final outcome. This approach is particularly useful in complex networks, where achieving the end goal may require multiple steps or gradual adaptation. For example, in a smart energy grid, instead of providing a single reward for peak-time energy conservation, a reward shaping approach might reward small reductions in energy use over time. By providing ongoing, incremental rewards, the mechanism encourages agents to adjust their behaviors gradually, which can be more effective for long-term adoption of energy-saving practices. Similarly, in public transit, reward shaping can encourage riders to travel during off-peak hours by providing incremental incentives for each off-peak journey, helping to reduce peak-time congestion sustainably. Reward shaping is also useful in managing social networks, where small rewards for behaviors like fact-checking, respectful interaction, or content verification can collectively lead to improved information quality and community health. By carefully structuring rewards, designers can guide agents toward behaviors that improve the system without needing to enforce strict penalties or rigid rules.

\subsection{Secure, Robust, and Resilient Design}

The secure, robust, and resilient design  aims at creating systems that can withstand and adapt to adversarial attacks, environmental disruptions, and operational failures \cite{zhu2020cross,rieger2019industrial,zhu2024disentangling}. In complex socio-technical networks, security, robustness, and resilience are essential for ensuring continuous operation and minimizing risks under adverse conditions. Designing these systems requires a structured approach to anticipate, manage, and recover from attacks, often by modeling potential threats and adversarial agents. By simulating these adversarial agents, designers can develop strategies that not only protect the system but also enable recovery and adaptation if an attack occurs.

Security in socio-technical systems focuses on protecting assets, data, and operational integrity from intentional threats. Security design involves anticipating adversarial agents with specific attack objectives, such as data theft, system sabotage, or physical harm. Adversarial agents are modeled based on their capabilities and intentions, allowing system designers to create defenses that counteract specific threats. Security measures may include encryption, intrusion detection, authentication, and access control, each tailored to thwart anticipated attacks. For example, an adversarial agent targeting data theft might be countered by encrypting sensitive information and implementing multi-factor authentication. A secure design framework prepares the system to withstand these targeted attacks, ensuring minimal disruption to functionality.

Robustness in secure design involves preparing the system to handle worst-case scenarios by considering zero-sum adversarial agents. In a zero-sum setup, an adversarial agent actively seeks to cause maximum harm, and robust designs anticipate and provide protection measures against such adversarial intrusions. Game-theoretic models often underpin robust design by setting up a competitive framework in which the designer and the adversary ``compete" for control or stability of the system. This allows the designer to consider and counteract the most damaging potential actions an adversary could take. For instance, in a power grid, robustness might involve ensuring that essential systems remain operational even under peak load attacks. The goal is to maintain stable operation despite aggressive interference, making the system inherently difficult to destabilize even under extreme adversarial conditions.

Resilience goes beyond preventing attacks to ensure that a system can adapt and recover if an attack succeeds. While robustness deals with pre-attack preparations, resilience focuses on post-attack responses, aiming to minimize long-term disruption and restore normal operations as quickly as possible \cite{rieger2013resilient,zhu2021control}. A resilient design framework includes proactive preparations—such as redundancy, failover protocols, and automated response strategies—that limit the impact of successful attacks. It also includes reactive strategies, like incident recovery processes and adaptive agent behaviors, to mitigate damage and maintain core functions. For instance, in a transportation network, resilience might involve rerouting traffic and activating backup systems if a cyberattack disables certain routes, maintaining the system’s overall operational integrity.

In the design of  secure, robust, and resilient systems, adversarial agents play a central role. Adversarial agents are modeled to represent potential attackers, enabling system designers to anticipate and counter specific tactics. Different types of adversarial agents are considered based on their goals and capabilities. For robustness design, adversarial agents are often modeled as zero-sum players, where the worst-case outcomes are analyzed and anticipated. These adversaries aim to cause maximum disruption or loss, requiring the system designer to consider all possible attack vectors. Robust design, therefore, involves anticipating the “strongest” adversarial tactics and implementing defensive strategies that minimize vulnerabilities. In security design, adversarial agents have specific attack objectives, such as stealing data or causing service outages. These agents may have different capabilities, ranging from passive observation to active manipulation. By understanding the likely methods of attack, designers can create security protocols that specifically counter these tactics. In resilience, adversarial agents are modeled as adaptive players who may escalate or change tactics in response to system defenses. This modeling is crucial in resilience design, as it enables the system to account for potential follow-up attacks or adaptations after an initial defense is activated.

Game theory provides a structured way to model interactions between the system designer and adversarial agents, facilitating the design of strategies that ensure security, robustness, and resilience \cite{zhu2024disentangling}. By framing the relationship as a game between defenders and adversaries, designers can anticipate adversarial tactics and optimize defensive responses accordingly. Security games enable the system designer to anticipate attacks and implement strategies to maximize the system’s protection \cite{AB11}. Robustness games focus on countering the worst-case tactics of adversarial agents, often framed as zero-sum games where the adversary seeks maximum disruption. Resilience games model adaptive adversarial behavior and the system’s responses, often involving multi-stage interactions where the system undergoes initial attacks, activates defense mechanisms, and then prepares for potential follow-up attacks.

Resilience requires both pre-attack preparation and post-attack adaptability. Pre-attack preparations often include redundancies, alert systems, and real-time monitoring that can detect and limit the impact of an initial attack. Post-attack strategies involve mechanisms that adaptively respond to the nature and scale of the disruption, aiming to restore normal operations or maintain critical functions. This two-tiered approach ensures that resilience strategies remain effective throughout the lifecycle of an attack. For instance, in a data center, pre-attack resilience may involve real-time data backups and automated threat detection, while post-attack resilience might activate failover systems and reconfigure compromised segments to maintain core functionality. By coordinating pre-attack and post-attack resilience strategies, system designers can ensure that socio-technical networks are not only resistant to attacks but also capable of rapid recovery and continued operation.

\section{Contemporary Applications}
\label{application}
Several key emerging application areas are crucial for next-generation control theorists. This section discusses several emerging application domains and the related role of game-theoretic control and design, including misinformation management in social networks, socio-technical control of infrastructure networks, and resilient design of socio-cyber-physical systems.

\subsection{Misinformation Management in Social Networks}

 Managing misinformation in social networks is complex due to their decentralized nature, diverse user motivations, and the viral spread of information. Game-theoretic design offers powerful tools to shape user behaviors, optimize information flows, and implement adaptive incentives that counter misinformation. By treating misinformation management as a strategic game among various players, including users, fact-checkers, platform operators, and adversaries—designers, can model behaviors, introduce incentives, and promote accuracy and reliability, \cite{amini2024control,li2023price,yang2023designing}.

\noindent{\bf Game-Theoretic Designs} Game-theoretic models analyze how misinformation spreads by assigning utility functions to each agent—whether an average user, a malicious actor, or the platform. For instance, signaling games can encourage users to prioritize credible information by attaching credibility indicators like verified labels. Misinformation management also involves adversarial and cooperative game strategies. Adversarial games help platforms deter malicious actors by implementing penalties, while cooperative games encourage users and fact-checkers to report false content through reward structures, leveraging collective effort to uphold information integrity.

Large networks benefit from mean-field game theory, which models the average user behavior rather than each individual, providing scalable solutions. By adjusting visibility algorithms for flagged content, platforms can limit misinformation spread without monitoring every user. Game-theoretic strategies also offer resilience, allowing platforms to anticipate adversarial tactics like coordinated campaigns or deepfakes. Resilience frameworks adapt in real time, adjusting detection criteria and response strategies during high-stakes periods such as elections.

\noindent{\bf Challenges} Managing misinformation is especially challenging due to the presence of adversarial agents who spread falsehoods intentionally. Game theory enables platforms to anticipate and counter these actions by modeling interactions as strategic games. In a zero-sum game setup, the platform's goal is to minimize misinformation spread while adversaries aim to maximize it, allowing designers to explore worst-case scenarios and implement countermeasures, like limiting the reach of unverified accounts in response to bot activity. Stackelberg games, modeling leader-follower dynamics, allow the platform to set policies that users respond to, such as penalties for misinformation, encouraging users to adjust their sharing behaviors accordingly. Multi-agent simulations provide insight into collective behaviors, allowing designers to test incentive structures and observe how misinformation strategies evolve.

Resilience requires adaptive responses that evolve with adversarial tactics. Game-theoretic models enable platforms to develop mechanisms that detect and counter misinformation in real time, adjusting criteria for detection and responses as new tactics like deepfakes emerge. By preparing for misinformation spikes during critical periods (e.g., elections), platforms can establish more robust detection protocols, increase fact-checking resources, and dynamically adjust incentives to favor verified content.

\subsection{Socio-Technical Control of Infrastructure Networks}

Infrastructure networks such as communication systems, smart grids, and transportation networks are essential for modern socio-technical ecosystems. Managing resource allocation and optimizing performance within these networks is challenging due to their scale, complexity, and the interdependence of agents--both human and machine--who interact within them. Game-theoretic design provides a structured approach to optimize the management of such networks by modeling interactions among agents, shaping behaviors, and strategically designing incentives that align individual actions with system-wide goals. By treating infrastructure networks as strategic environments, game theory allows system designers to anticipate behaviors, manage congestion, allocate resources efficiently, and ensure resilience, even in dynamic and unpredictable conditions \cite{chen2019dynamic,chen2016interdependent,huang2017large,huang2018factored}.

\noindent{\bf Game-Theoretic Designs} Game theory offers several models applicable to the control of infrastructure networks, each addressing different aspects of agent interaction and resource allocation \cite{EB18}. The central idea is to treat each agent—whether a vehicle, a consumer, or a data packet—as a rational player with individual objectives. The system designer’s role is to craft rules, incentives, and mechanisms that encourage agents to make decisions that support overall network performance. Congestion games, for instance, model scenarios where agents compete for limited resources that become less efficient as usage increases, such as bandwidth in communication networks or lanes in transportation systems. In these games, each agent’s payoff decreases as more agents choose the same resource, creating a natural disincentive for overcrowding. By structuring incentives, such as variable pricing for bandwidth, congestion games encourage agents to spread out across resources, minimizing bottlenecks and improving network efficiency. In large-scale networks, managing individual agent behaviors is often impractical due to the sheer number of participants. Mean-field game theory provides a viable approach by focusing on the average or aggregate behavior of agents rather than tracking each one individually. Mean-field games are particularly useful in infrastructure networks like smart grids, where thousands or even millions of consumers interact with the grid simultaneously. By modeling the collective behavior of users, mean-field theory enables designers to implement policies that affect overall consumption trends.



\noindent{\bf Challenges} Infrastructure networks often feature hierarchical structures where certain agents, such as network operators or grid managers, act as leaders who set policies or pricing strategies, while others, like users or vehicles, respond to these decisions. Stackelberg games effectively model this leader-follower dynamic, where leaders optimize their strategies by anticipating follower responses. For example, a utility company may adjust electricity prices based on projected demand, encouraging consumers to modify their usage patterns, which helps balance grid demand and prevent overloads. In this way, Stackelberg models create a feedback loop where the leader’s decisions shape agent behaviors, which in turn stabilize the overall system.

These networks are also inherently dynamic, with agents continuously adapting their strategies based on observed outcomes and interactions. Dynamic game theory captures this adaptive nature, modeling how agents adjust their actions over time to optimize their objectives in response to others. 
 Resilience is equally crucial in infrastructure networks, requiring the ability to adapt to evolving conditions and recover from disruptions. Game-theoretic resilience modeling anticipates the responses of various agents to unexpected events, such as power outages or natural disasters, and helps design strategic responses that maintain system integrity. During a grid disruption, for instance, game-theoretic models can guide the allocation of energy to critical areas while incentivizing non-essential users to reduce consumption temporarily. Similarly, in transportation networks, game-theoretic resilience models support real-time route adjustments during emergencies, signaling drivers to use alternative paths and helping prevent gridlock. Resilient infrastructure design, informed by game-theoretic insights, creates adaptive, flexible systems capable of sustaining core functions while enabling agents to respond effectively to incentives and disruptions. By aligning individual agent behavior with system-wide resilience goals, these models facilitate continuous, reliable performance, even under challenging conditions.

\subsection{Resilience in Socio-Cyber-Physical Systems through Game-Theoretic Design}

Socio-cyber-physical systems (SCPS), encompassing infrastructures such as smart cities, autonomous transportation networks, energy grids, and water management systems, combine physical infrastructure, digital control, and human elements. In SCPS, resilience is essential, as these systems must withstand and adapt to a range of threats, from cyber-physical attacks and natural disruptions to human-driven errors and social manipulation. Humans, as part of these systems, can introduce new vulnerabilities, sometimes serving as the weakest link due to unintentional errors, misjudgments, or even malicious actions. Game-theoretic design offers a powerful framework to enhance resilience in SCPS by modeling both adversarial and human interactions, optimizing response strategies, and creating incentive structures that mitigate human vulnerabilities. By treating resilience as a strategic game involving cyber, physical, and social elements, game-theoretic approaches allow SCPS to anticipate, counteract, and recover from a spectrum of disruptions, ensuring stability and performance \cite{zhu2024disentangling,huang2020dynamic,zhao2022multi}.

\noindent{\bf Game-Theoretic Models} In resilient SCPS,  game-theoretic frameworks can model interactions between the system and adversarial agents seeking to maximize disruption. By considering adversarial scenarios, this approach prepares the system to handle rationalized actions from adversarial agents and accidental human errors.   Dynamic game theory plays a critical role in SCPS because it allows the system to adapt strategies based on changing threat landscapes and evolving human behaviors. In dynamic games, agents adjust their strategies according to observed outcomes, enabling the system to evolve toward greater resilience against both technical threats and human vulnerabilities. For example, in an industrial control system, adversarial agents may target network communications, while human operators may inadvertently disrupt operations through procedural mistakes. An dynamic game framework lets the system refine its defense mechanisms and response protocols based on these incidents, adapting to new threats and changing human behaviors.  

Real-time response is crucial for SCPS resilience, as delays in counteracting disruptions can lead to cascading failures. Dynamic games enable the system to respond to threats and errors in real time, continuously updating its strategies based on ongoing interactions with adversarial and human agents. For example, in a water distribution network, the control system might respond immediately to sabotage or accidental contamination, isolating compromised sections and redirecting resources to unaffected areas. A dynamic game framework allows SCPS to adapt quickly, maintaining core functions by adjusting to new information about cyber, physical, or social threats, including human errors or misjudgments. This real-time adaptability enhances SCPS resilience by mitigating the immediate impact of both malicious actions and human mistakes.

\noindent{\bf Challenges} Resilience in socio-cyber-physical systems (SCPS) involves not only enduring attacks and disruptions but also ensuring effective recovery and adaptation afterward. A key challenge lies in developing SCPS that can quickly recover following disruptions, whether they arise from technical failures or social disturbances. Game-theoretic design aids in this recovery process by enabling SCPS to dynamically adjust strategies based on technical and social impacts. Game-theoretic models simulate post-disruption scenarios, helping to identify optimal recovery steps. For example, in a water distribution network, an SCPS can prioritize the reactivation of critical pipelines while isolating compromised areas, ensuring both service continuity and public trust.

Adaptive learning mechanisms, which evolve based on past disruptions, are critical for maintaining resilience in SCPS. Multi-agent learning theory offers a framework that allows SCPS to learn from incidents, refining defense strategies to mitigate vulnerability to similar future attacks. Social learning is equally important, as it enables the system to monitor and integrate human responses into its resilience approach. Following a significant cyber-attack on a smart grid, for instance, the system can assess which defenses were most effective and how users reacted, using these insights to adjust protocols and increase resilience. This continuous adaptation through multi-agent learning and dynamic game theory enable SCPS to address both human and technical vulnerabilities.

Human factors pose unique challenges to SCPS resilience, as humans often represent the weakest link. To anticipate and manage human errors and vulnerabilities, it is essential to incorporate behavioral biases and tendencies into SCPS decision models. By accounting for these human factors and implementing robust failover capabilities, SCPS can better prepare for human errors, mitigate the risks they pose, and strengthen overall resilience. This comprehensive approach, integrating technical resilience with an understanding of human vulnerabilities, is vital for robust and adaptive SCPS.

\section{Conclusions}
\label{conclusion}
The development of resilient and secure socio-technical networks and SCPS requires sophisticated game-theoretic frameworks that manage the complex interplay between human, cyber, and physical elements. Successfully addressing the challenges of resilience, misinformation management, and infrastructure optimization requires balancing decentralized incentives with system-wide stability and security. Frameworks such as Stackelberg games for leader-follower dynamics, dynamic mechanism design for adaptive incentives, and multi-layered rewards and penalties allow network operators to optimize performance, reinforce resilience, and deter adversarial behaviors. By incorporating adaptive learning and real-time response mechanisms, these models enable SCPS to withstand evolving threats, such as cyber-attacks and misinformation, while maintaining  operations across diverse conditions. Future research should focus on expanding game-theoretic tools to address scalability and predictive capabilities, ultimately creating socio-technical systems that are not only robust but also adaptable and secure in the face of emerging, complex challenges.

In this letter our focus has been primarily on hierarchical decision making with two levels of hierarchy (a leader or leaders at the top and a follower or followers at the bottom), where naturally the mode of decision-making is asymmetric across leader(s) and follower(s). It is also possible to introduce intermediate layers in the decision-making process, with the Stackelberg game solution extended to such settings; see, e.g., \cite{TB83,TB89}. What has also not been covered here is the symmetric mode of decision making, that is with only one layer where agents sit, in which case within the noncooperative setting Nash equilibrium would be the appropriate solution concept to adopt; for details see \cite{bacsar1998dynamic,TB89b}. 

\bibliographystyle{IEEEtran1}
\bibliography{IEEEabrv_LCSS_2024,LCSS2024Opinion}

\begin{thebibliography}{100}
\providecommand{\url}[1]{#1}
\csname url@samestyle\endcsname
\providecommand{\newblock}{\relax}
\providecommand{\bibinfo}[2]{#2}
\providecommand{\BIBentrySTDinterwordspacing}{\spaceskip=0pt\relax}
\providecommand{\BIBentryALTinterwordstretchfactor}{4}
\providecommand{\BIBentryALTinterwordspacing}{\spaceskip=\fontdimen2\font plus
\BIBentryALTinterwordstretchfactor\fontdimen3\font minus
  \fontdimen4\font\relax}
\providecommand{\BIBforeignlanguage}[2]{{%
\expandafter\ifx\csname l@#1\endcsname\relax
\typeout{** WARNING: IEEEtran.bst: No hyphenation pattern has been}%
\typeout{** loaded for the language `#1'. Using the pattern for}%
\typeout{** the default language instead.}%
\else
\language=\csname l@#1\endcsname
\fi
#2}}
\providecommand{\BIBdecl}{\relax}
\BIBdecl

\bibitem{vonNeumann1944}
J.~von Neumann and O.~Morgenstern, \emph{Theory of Games and Economic
  Behavior}.\hskip 1em plus 0.5em minus 0.4em\relax Princeton University Press,
  1944.

\bibitem{nash1950}
J.~Nash, ``Equilibrium points in {N-person} games,'' \emph{Proceedings of the
  National Academy of Sciences}, vol.~36, no.~1, pp. 48--49, 1950.

\bibitem{nash1951}
J.~Nash, ``Non-cooperative games,'' \emph{Annals of Mathematics}, vol.~54,
  no.~2, pp. 286--295, 1951.

\bibitem{von2010market}
H.~{Von Stackelberg}, \emph{Market Structure and Equilibrium}.\hskip 1em plus
  0.5em minus 0.4em\relax Springer Science \& Business Media, 2010.

\bibitem{bellman1957}
R.~Bellman, \emph{Dynamic Programming}.\hskip 1em plus 0.5em minus 0.4em\relax
  Princeton University Press, 1957.

\bibitem{isaacs1965}
R.~Isaacs, \emph{Differential Games: A Mathematical Theory with Applications to
  Warfare and Pursuit, Control and Optimization}.\hskip 1em plus 0.5em minus
  0.4em\relax Wiley, 1965.

\bibitem{fudenberg1991}
D.~Fudenberg and J.~Tirole, \emph{Game Theory}.\hskip 1em plus 0.5em minus
  0.4em\relax MIT Press, 1991.

\bibitem{helbing2013}
D.~Helbing, ``Globally networked risks and how to respond,'' \emph{Nature},
  vol. 497, pp. 51--59, 2013.

\bibitem{zimmerman2017conceptual}
R.~Zimmerman, Q.~Zhu, F.~de~Leon, and Z.~Guo, ``Conceptual modeling framework
  to integrate resilient and interdependent infrastructure in extreme
  weather,'' \emph{Journal of Infrastructure Systems}, vol.~23, no.~4, p.
  04017034, 2017.

\bibitem{zimmerman2016promoting}
R.~Zimmerman, Q.~Zhu, and C.~Dimitri, ``Promoting resilience for food, energy,
  and water interdependencies,'' \emph{Journal of Environmental Studies and
  Sciences}, vol.~6, pp. 50--61, 2016.

\bibitem{braess1968}
D.~Braess, ``Über ein paradoxon aus der verkehrsplanung,''
  \emph{Unternehmensforschung}, vol.~12, pp. 258--268, 1968.

\bibitem{roughgarden2005}
T.~Roughgarden, \emph{Selfish Routing and the Price of Anarchy}.\hskip 1em plus
  0.5em minus 0.4em\relax MIT Press, 2005.

\bibitem{saad2012}
W.~Saad, Z.~Han, H.~V. Poor, and T.~Ba\c{s}ar, ``Game-theoretic methods for the
  smart grid: An overview of microgrid systems, demand-side management, and
  smart grid communications,'' \emph{IEEE Signal Processing Magazine}, vol.~29,
  no.~5, pp. 86--105, 2012.

\bibitem{stoustrup2019smart}
J.~Stoustrup, A.~Annaswamy, A.~Chakrabortty, and Z.~Qu, ``Smart grid control,''
  \emph{Springer International Publishing. doi}, vol.~10, pp. 978--3, 2019.

\bibitem{liu2022herd}
S.~Liu, Y.~Zhao, and Q.~Zhu, ``Herd behaviors in epidemics: A dynamics-coupled
  evolutionary games approach,'' \emph{Dynamic Games and Applications},
  vol.~12, no.~1, pp. 183--213, 2022.

\bibitem{PBB20}
P.~Pare, C.~Beck, and T.~Ba\c{s}ar, ``Modeling, estimation, and analysis of
  epidemics over networks: An overview,'' \emph{Annual Reviews in Control},
  vol.~50, pp. 345--360, 2020.

\bibitem{gubar2018optimal}
E.~Gubar, V.~Taynitskiy, and Q.~Zhu, ``Optimal control of heterogeneous
  mutating viruses,'' \emph{Games}, vol.~9, no.~4, p. 103, 2018.

\bibitem{hayel2017epidemic}
Y.~Hayel and Q.~Zhu, ``Epidemic protection over heterogeneous networks using
  evolutionary poisson games,'' \emph{IEEE Transactions on Information
  Forensics and Security}, vol.~12, no.~8, pp. 1786--1800, 2017.

\bibitem{zhu2022preface}
Q.~Zhu, E.~Gubar, and E.~Altman, ``Preface to special issue on dynamic games
  for modeling and control of epidemics,'' \emph{Dynamic Games and
  Applications}, vol.~12, no.~1, pp. 1--6, 2022.

\bibitem{myerson1991}
R.~B. Myerson, \emph{Game Theory: Analysis of Conflict}.\hskip 1em plus 0.5em
  minus 0.4em\relax Harvard University Press, 1991.

\bibitem{goyal2023networks}
S.~Goyal, \emph{Networks: An Economics Approach}.\hskip 1em plus 0.5em minus
  0.4em\relax MIT Press, 2023.

\bibitem{jackson2008social}
M.~Jackson, \emph{Social and Economic Networks}.\hskip 1em plus 0.5em minus
  0.4em\relax Princeton University Press, 2008.

\bibitem{zhao2024learning}
Y.~Zhao, ``Learning and game-theoretic paradigms for strategic coordination of
  multi-agent autonomous systems,'' Ph.D. dissertation, New York University
  Tandon School of Engineering, 2024.

\bibitem{li2022role}
T.~Li, Y.~Zhao, and Q.~Zhu, ``The role of information structures in
  game-theoretic multi-agent learning,'' \emph{Annual Reviews in Control},
  vol.~53, pp. 296--314, 2022.

\bibitem{huang2020dynamic}
Y.~Huang, J.~Chen, L.~Huang, and Q.~Zhu, ``Dynamic games for secure and
  resilient control system design,'' \emph{National Science Review}, vol.~7,
  no.~7, pp. 1125--1141, 2020.

\bibitem{bacsar1998dynamic}
T.~Ba{\c{s}}ar and G.~J. Olsder, \emph{Dynamic Noncooperative Game
  Theory}.\hskip 1em plus 0.5em minus 0.4em\relax SIAM, 1998.

\bibitem{BZ18}
T.~Ba\c{s}ar and G.~Zaccour, Eds., \emph{Handbook of Dynamic Game Theory,
  Volume I (Theory of Dynamic Games), Volume II (Applications of Dynamic
  Games)}.\hskip 1em plus 0.5em minus 0.4em\relax Springer International
  Publishing, 2018.

\bibitem{zhu2015game}
Q.~Zhu and T.~Ba\c{s}ar, ``Game-theoretic methods for robustness, security, and
  resilience of cyberphysical control systems: games-in-games principle for
  optimal cross-layer resilient control systems,'' \emph{IEEE Control Systems
  Magazine}, vol.~35, no.~1, pp. 46--65, 2015.

\bibitem{zhu2024disentangling}
Q.~Zhu and T.~Ba{\c{s}}ar, ``Disentangling resilience from robustness:
  Contextual dualism, interactionism, and game-theoretic paradigms,''
  \emph{IEEE Control Systems Magazine}, vol.~44, no.~3, pp. 95--103, 2024.

\bibitem{ponte2016holism}
B.~Ponte, J.~Costas, J.~Puche, D.~De~la Fuente, and R.~Pino, ``Holism versus
  reductionism in supply chain management: An economic analysis,''
  \emph{Decision Support Systems}, vol.~86, pp. 83--94, 2016.

\bibitem{verschuuren2017holism}
G.~M. Verschuuren, \emph{The Holism-Reductionism Debate: In Physics, Genetics,
  Biology, Neuroscience, Ecology, and Sociology}.\hskip 1em plus 0.5em minus
  0.4em\relax CreateSpace, 2017.

\bibitem{chen2019game}
J.~Chen and Q.~Zhu, \emph{A Game-and Decision-Theoretic Approach to Resilient
  Interdependent Network Analysis and Design}.\hskip 1em plus 0.5em minus
  0.4em\relax Springer, 2019.

\bibitem{albaba2019modeling}
B.~M. Albaba and Y.~Yildiz, ``Modeling cyber-physical human systems via an
  interplay between reinforcement learning and game theory,'' \emph{Annual
  Reviews in Control}, vol.~48, pp. 1--21, 2019.

\bibitem{bacsar1986tutorial}
T.~Ba{\c{s}}ar, ``A tutorial on dynamic and differential games,'' in
  \emph{Dynamic Games and Applications in Economics}, ser. Lecture Notes in
  Economics and Mathematical Systems.\hskip 1em plus 0.5em minus 0.4em\relax
  Springer, 1986, vol. 265, pp. 1--25.

\bibitem{basear1984feedback}
T.~Ba\c{s}ar and A.~Haurie, ``Feedback equilibria in differential games with
  structural and modal uncertainties, vol. 1 of advances in large scale
  systems,'' \emph{Connecticut: JAI Press Inc., JB Cruz, Jr., editor}, pp.
  163--201, 1984.

\bibitem{zhao2022stackelberg}
Y.~Zhao, B.~Huang, J.~Yu, and Q.~Zhu, ``Stackelberg strategic guidance for
  heterogeneous robots collaboration,'' in \emph{2022 International Conference
  on Robotics and Automation (ICRA)}.\hskip 1em plus 0.5em minus 0.4em\relax
  IEEE, 2022, pp. 4922--4928.

\bibitem{zhao2023stackelberg}
Y.~Zhao and Q.~Zhu, ``Stackelberg meta-learning for strategic guidance in
  multi-robot trajectory planning,'' in \emph{2023 IEEE/RSJ International
  Conference on Intelligent Robots and Systems (IROS)}.\hskip 1em plus 0.5em
  minus 0.4em\relax IEEE, 2023, pp. 11\,342--11\,347.

\bibitem{maharjan2013dependable}
S.~Maharjan, Q.~Zhu, Y.~Zhang, S.~Gjessing, and T.~Ba\c{s}ar, ``Dependable
  demand response management in the smart grid: A {Stackelberg} game
  approach,'' \emph{IEEE Transactions on Smart Grid}, vol.~4, no.~1, pp.
  120--132, 2013.

\bibitem{chen2017stackelberg}
J.~Chen and Q.~Zhu, ``A {Stackelberg} game approach for two-level distributed
  energy management in smart grids,'' \emph{IEEE Transactions on Smart Grid},
  vol.~9, no.~6, pp. 6554--6565, 2017.

\bibitem{GBD-G16}
B.~Gharesifard, T.~Ba\c{s}ar, and A.~D. Dominguez-Garcia, ``Price-based
  coordinated aggregation of networked distributed energy resources,''
  \emph{IEEE Transactions on Automatic Control}, vol.~61, no.~10, pp.
  2936--2946, October 2016.

\bibitem{ALCB21}
K.~Alshehri, J.~Liu, X.~Chen, and T.~Ba\c{s}ar, ``A game-theoretic framework
  for multi-period multi-company demand response management in the smart
  grid,'' \emph{IEEE Transactions on Control Systems Technology}, vol.~29,
  no.~3, pp. 1019--1034, May 2021.

\bibitem{yang2007stackelberg}
H.~Yang, X.~Zhang, and Q.~Meng, ``Stackelberg games and multiple equilibrium
  behaviors on networks,'' \emph{Transportation Research Part B:
  Methodological}, vol.~41, no.~8, pp. 841--861, 2007.

\bibitem{korilis1997achieving}
Y.~A. Korilis, A.~A. Lazar, and A.~Orda, ``Achieving network optima using
  {Stackelberg} routing strategies,'' \emph{IEEE/ACM Transactions on
  Networking}, vol.~5, no.~1, pp. 161--173, 1997.

\bibitem{pan2022poisoned}
Y.~Pan and Q.~Zhu, ``On poisoned {Wardrop} equilibrium in congestion games,''
  in \emph{International Conference on Decision and Game Theory for
  Security}.\hskip 1em plus 0.5em minus 0.4em\relax Springer, 2022, pp.
  191--211.

\bibitem{pawlick2016stackelberg}
J.~Pawlick and Q.~Zhu, ``A {Stackelberg} game perspective on the conflict
  between machine learning and data obfuscation,'' in \emph{2016 IEEE
  International Workshop on Information Forensics and Security (WIFS)}.\hskip
  1em plus 0.5em minus 0.4em\relax IEEE, 2016, pp. 1--6.

\bibitem{maschler2013game}
M.~Maschler, E.~Solan, and S.~Zamir, \emph{Game Theory}.\hskip 1em plus 0.5em
  minus 0.4em\relax Cambridge University Press, 2013.

\bibitem{chen2023qos}
J.~Chen, J.~Farooq, and Q.~Zhu, ``{QoS}-based contract design for profit
  maximization in {IoT}-enabled data markets,'' \emph{IEEE Internet of Things
  Journal}, vol.~10, no.~11, pp. 10\,080--10\,094, 2023.

\bibitem{farooq2018optimal}
M.~J. Farooq and Q.~Zhu, ``Optimal dynamic contract for spectrum reservation in
  mission-critical {UNB-IoT} systems,'' in \emph{2018 16th International
  Symposium on Modeling and Optimization in Mobile, Ad Hoc, and Wireless
  Networks (WiOpt)}.\hskip 1em plus 0.5em minus 0.4em\relax IEEE, 2018, pp.
  1--6.

\bibitem{zhu2008state}
Q.~Zhu and L.~Pavel, ``State-space approach to pricing design in {OSNR} {Nash}
  game,'' \emph{IFAC Proceedings}, vol.~41, no.~2, pp. 12\,001--12\,006, 2008.

\bibitem{chen2016optimal}
J.~Chen and Q.~Zhu, ``Optimal contract design under asymmetric information for
  cloud-enabled internet of controlled things,'' in \emph{International
  Conference on Decision and Game Theory for Security}.\hskip 1em plus 0.5em
  minus 0.4em\relax Springer, 2016, pp. 329--348.

\bibitem{chen2022transactive}
J.~Chen, Y.~Huang, and Q.~Zhu, ``Transactive resilience in renewable
  microgrids: A contract-theoretic approach,'' in \emph{2022 56th Annual
  Conference on Information Sciences and Systems (CISS)}.\hskip 1em plus 0.5em
  minus 0.4em\relax IEEE, 2022, pp. 1--6.

\bibitem{chen2016game}
J.~Chen and Q.~Zhu, ``A game-theoretic framework for resilient and distributed
  generation control of renewable energies in microgrids,'' \emph{IEEE
  Transactions on Smart Grid}, vol.~8, no.~1, pp. 285--295, 2016.

\bibitem{zhu2012game}
Q.~Zhu, J.~Zhang, P.~W. Sauer, A.~Dominguez-Garcia, and T.~Ba{\c{s}}ar, ``A
  game-theoretic framework for control of distributed renewable-based energy
  resources in smart grids,'' in \emph{2012 American Control Conference
  (ACC)}.\hskip 1em plus 0.5em minus 0.4em\relax IEEE, 2012, pp. 3623--3628.

\bibitem{ma2022spatio}
H.~Ma, F.~Fang, and D.~C. Parkes, ``Spatio-temporal pricing for ridesharing
  platforms,'' \emph{Operations Research}, vol.~70, no.~2, pp. 1025--1041,
  2022.

\bibitem{hong2022optimal}
J.~H. Hong and X.~Liu, ``The optimal pricing for green ride services in the
  ride-sharing economy,'' \emph{Transportation Research Part D: Transport and
  Environment}, vol. 104, p. 103205, 2022.

\bibitem{nagurney2010sustainable}
A.~Nagurney and L.~S. Nagurney, ``Sustainable supply chain network design: A
  multicriteria perspective,'' \emph{International Journal of Sustainable
  Engineering}, vol.~3, no.~3, pp. 189--197, 2010.

\bibitem{leng2005game}
M.~Leng and M.~Parlar, ``Game theoretic applications in supply chain
  management: A review,'' \emph{INFOR: Information Systems and Operational
  Research}, vol.~43, no.~3, pp. 187--220, 2005.

\bibitem{gale1962college}
D.~Gale and L.~S. Shapley, ``College admissions and the stability of
  marriage,'' \emph{The American Mathematical Monthly}, vol.~69, no.~1, pp.
  9--15, 1962.

\bibitem{jackson2002coalition}
M.~O. Jackson and B.~Moselle, ``Coalition and party formation in a legislative
  voting game,'' \emph{Journal of Economic Theory}, vol. 103, no.~1, pp.
  49--87, 2002.

\bibitem{lombardi2021model}
C.~Lombardi, L.~Picado-Santos, and A.~M. Annaswamy, ``Model-based dynamic toll
  pricing: An overview,'' \emph{Applied Sciences}, vol.~11, no.~11, p. 4778,
  2021.

\bibitem{SB07}
H.~Shen and T.~Ba\c{s}ar, ``Optimal nonlinear pricing for a monopolistic
  network service provider with complete and incomplete information,''
  \emph{IEEE J. on Selected Areas in Communications (JSAC) Special Issue:
  Non-Cooperative Behavior in Networking}, vol.~25, no.~6, pp. 1216--1223, June
  2007.

\bibitem{SB11}
H.~Shen and T.~Ba\c{s}ar, ``Pricing under information asymmetry for a large
  population of users,'' \emph{Telecommunication Systems}, vol.~47, no. 1-2,
  pp. 123--136, June 2011.

\bibitem{yang2023designing}
Y.-T. Yang, T.~Li, and Q.~Zhu, ``Designing policies for truth: Combating
  misinformation with transparency and information design,'' in \emph{2023 21st
  International Symposium on Modeling and Optimization in Mobile, Ad Hoc, and
  Wireless Networks (WiOpt)}.\hskip 1em plus 0.5em minus 0.4em\relax IEEE,
  2023, pp. 127--134.

\bibitem{zhang2019optimal}
T.~Zhang and Q.~Zhu, ``Optimal two-sided market mechanism design for
  large-scale data sharing and trading in massive {IoT} networks,'' \emph{arXiv
  preprint arXiv:1912.06229}, 2019.

\bibitem{zhu2012differential}
Q.~Zhu, Z.~Han, and T.~Ba{\c{s}}ar, ``A differential game approach to
  distributed demand side management in smart grid,'' in \emph{2012 IEEE
  International Conference on Communications (ICC)}.\hskip 1em plus 0.5em minus
  0.4em\relax IEEE, 2012, pp. 3345--3350.

\bibitem{zhang2023dynamic}
T.~Zhang, ``Dynamic mechanism design: From theories to applications,'' Ph.D.
  dissertation, New York University Tandon School of Engineering, 2023.

\bibitem{chen2021dynamic}
J.~Chen, Q.~Zhu, and T.~Ba{\c{s}}ar, ``Dynamic contract design for systemic
  cyber risk management of interdependent enterprise networks,'' \emph{Dynamic
  Games and Applications}, vol.~11, no.~2, pp. 294--325, 2021.

\bibitem{zhang2022incentive}
T.~Zhang and Q.~Zhu, ``On incentive compatibility in dynamic mechanism design
  with exit option in a {Markovian} environment,'' \emph{Dynamic Games and
  Applications}, vol.~12, no.~2, pp. 701--745, 2022.

\bibitem{nagurney2021braess}
A.~Nagurney and L.~S. Nagurney, ``The {Braess} paradox,'' in
  \emph{International Encyclopedia of Transportation}, R.~Vickerman, Ed.\hskip
  1em plus 0.5em minus 0.4em\relax Oxford: Elsevier, 2021, pp. 601--607.

\bibitem{yang2024prada}
Y.-T. Yang, H.~Lei, and Q.~Zhu, ``{PRADA}: Proactive risk assessment and
  mitigation of misinformed demand attacks on navigational route
  recommendations,'' \emph{arXiv preprint arXiv:2409.00243}, 2024.

\bibitem{yang2023strategic}
Y.-T. Yang, H.~Lei, and Q.~Zhu, ``Strategic information attacks on
  incentive-compatible navigational recommendations in intelligent
  transportation systems,'' \emph{arXiv preprint arXiv:2310.01646}, 2023.

\bibitem{altman2006survey}
E.~Altman, T.~Boulogne, R.~El-Azouzi, T.~Jim{\'e}nez, and L.~Wynter, ``A survey
  on networking games in telecommunications,'' \emph{Computers \& Operations
  Research}, vol.~33, no.~2, pp. 286--311, 2006.

\bibitem{altman2001avoiding}
E.~Altman, R.~El~Azouzi, and O.~Pourtallier, ``Avoiding paradoxes in routing
  games,'' in \emph{Teletraffic Science and Engineering}.\hskip 1em plus 0.5em
  minus 0.4em\relax Elsevier, 2001, vol.~4, pp. 643--654.

\bibitem{simon1990bounded}
H.~A. Simon, ``Bounded rationality,'' \emph{Utility and Probability}, pp.
  15--18, 1990.

\bibitem{huang2023review}
L.~Huang and Q.~Zhu, ``Review of system-scientific perspectives for analysis,
  exploitation, and mitigation of cognitive vulnerabilities,'' in
  \emph{Cognitive Security: A System-Scientific Approach}.\hskip 1em plus 0.5em
  minus 0.4em\relax Springer, 2023, pp. 49--65.

\bibitem{sims2003implications}
C.~A. Sims, ``Implications of rational inattention,'' \emph{Journal of Monetary
  Economics}, vol.~50, no.~3, pp. 665--690, 2003.

\bibitem{kahneman2013prospect}
D.~Kahneman and A.~Tversky, ``Prospect theory: An analysis of decision under
  risk,'' in \emph{Handbook of the Fundamentals of Financial Decision Making:
  Part I}.\hskip 1em plus 0.5em minus 0.4em\relax World Scientific, 2013, pp.
  99--127.

\bibitem{SSB20}
A.~Sanjab, W.~Saad, and T.~Ba\c{s}ar, ``A game of drones: Cyber-physical
  security of time-critical {UAV} applications with cumulative prospect theory
  perceptions and valuations,'' \emph{IEEE Transactions on Communications},
  vol.~68, no.~11, pp. 6990--7006, November 2020.

\bibitem{TB21}
T.~Ba\c{s}ar, ``Robust designs through risk sensitivity: An overview,''
  \emph{J. Systems Science and Complexity}, vol.~34, pp. 1634--1665, October
  2021.

\bibitem{huang2023cognitive}
L.~Huang and Q.~Zhu, \emph{Cognitive Security: A System-Scientific
  Approach}.\hskip 1em plus 0.5em minus 0.4em\relax Springer Nature, 2023.

\bibitem{huang2024psyborg+}
S.~Huang and Q.~Zhu, ``{PsybORG$^+$}: Cognitive modeling for triggering and
  detection of cognitive biases of advanced persistent threats,'' \emph{arXiv
  preprint arXiv:2408.01310}, 2024.

\bibitem{huang2022radams}
L.~Huang and Q.~Zhu, ``{RADAMS}: Resilient and adaptive alert and attention
  management strategy against informational denial-of-service {(IDoS)}
  attacks,'' \emph{Computers \& Security}, vol. 121, p. 102844, 2022.

\bibitem{basar2020recent}
T.~Ba\c{s}ar, ``Recent advances in mean field games,'' \emph{Proc. 7th
  International Conference on Control and Optimization with Industrial
  Applications}, vol.~1, p.~14, 2020.

\bibitem{caines2021mean}
P.~E. Caines, ``Mean field games,'' in \emph{Encyclopedia of Systems and
  Control}.\hskip 1em plus 0.5em minus 0.4em\relax Springer, 2021, pp.
  1197--1202.

\bibitem{li2022confluence}
T.~Li, G.~Peng, Q.~Zhu, and T.~Ba{\c{s}}ar, ``The confluence of networks,
  games, and learning a game-theoretic framework for multiagent decision making
  over networks,'' \emph{IEEE Control Systems Magazine}, vol.~42, no.~4, pp.
  35--67, 2022.

\bibitem{ZYB21}
K.~Zhang, Z.~Yang, and T.~Ba\c{s}ar, ``Multi-agent reinforcement learning: A
  selective overview of theories and algorithms,'' in \emph{Handbook of
  Reinforcement Learning and Control}, ser. Studies in Systems, Decision and
  Control 325, K.~G.~V. et~al., Ed.\hskip 1em plus 0.5em minus 0.4em\relax
  Springer Nature Switzerland AG, 2021, pp. 321--384.

\bibitem{kairouz2021advances}
P.~Kairouz, H.~B. McMahan, B.~Avent, A.~Bellet, M.~Bennis, A.~N. Bhagoji,
  K.~Bonawitz, Z.~Charles, G.~Cormode, R.~Cummings \emph{et~al.}, ``Advances
  and open problems in federated learning,'' \emph{Foundations and
  Trends{\textregistered} in Machine Learning}, vol.~14, no. 1--2, pp. 1--210,
  2021.

\bibitem{zhang2021survey}
C.~Zhang, Y.~Xie, H.~Bai, B.~Yu, W.~Li, and Y.~Gao, ``A survey on federated
  learning,'' \emph{Knowledge-Based Systems}, vol. 216, p. 106775, 2021.

\bibitem{li2023self}
T.~Li, H.~Lei, and Q.~Zhu, ``Self-adaptive driving in nonstationary
  environments through conjectural online lookahead adaptation,'' in \emph{2023
  IEEE International Conference on Robotics and Automation (ICRA)}.\hskip 1em
  plus 0.5em minus 0.4em\relax IEEE, 2023, pp. 7205--7211.

\bibitem{li2024multi}
T.~Li, Z.~Bian, H.~Lei, F.~Zuo, Y.-T. Yang, Q.~Zhu, Z.~Li, and K.~Ozbay,
  ``Multi-level traffic-responsive tilt camera surveillance through predictive
  correlated online learning,'' \emph{Transportation Research Part C: Emerging
  Technologies}, vol. 167, p. 104804, 2024.

\bibitem{ishii2022security}
H.~Ishii and Q.~Zhu, \emph{Security and Resilience of Control Systems}.\hskip
  1em plus 0.5em minus 0.4em\relax Springer, 2022.

\bibitem{ben-tal2009robust}
A.~Ben-Tal, L.~E. Ghaoui, and A.~Nemirovski, \emph{Robust Optimization}, ser.
  Princeton Series in Applied Mathematics.\hskip 1em plus 0.5em minus
  0.4em\relax Princeton University Press, 2009, published on August 30, 2009.

\bibitem{bacsar2008h}
T.~Ba{\c{s}}ar and P.~Bernhard, \emph{H-infinity Optimal Control and Related
  Minimax Design Problems: A Dynamic Game Approach}.\hskip 1em plus 0.5em minus
  0.4em\relax Springer Science \& Business Media, 2008.

\bibitem{zhu2024foundations}
Q.~Zhu, ``Foundations of cyber resilience: The confluence of game, control, and
  learning theories,'' \emph{arXiv preprint arXiv:2404.01205}, 2024.

\bibitem{rass2018game}
S.~Rass and S.~Schauer, \emph{Game Theory for Security and Risk
  Management}.\hskip 1em plus 0.5em minus 0.4em\relax Springer, 2018, vol.~10.

\bibitem{chen2019interdependent}
J.~Chen and Q.~Zhu, ``Interdependent strategic security risk management with
  bounded rationality in the internet of things,'' \emph{IEEE Transactions on
  Information Forensics and Security}, vol.~14, no.~11, pp. 2958--2971, 2019.

\bibitem{bergemann2019information}
D.~Bergemann and S.~Morris, ``Information design: A unified perspective,''
  \emph{Journal of Economic Literature}, vol.~57, no.~1, pp. 44--95, 2019.

\bibitem{TB24}
T.~Ba\c{s}ar, ``Inducement of desired behavior via soft policies,''
  \emph{International Game Theory Review, Special Issue on Game theory and
  Optimization}, vol.~26, no.~2, p. 2440002 (26 pp), June 2024.

\bibitem{li2024decision}
T.~Li, Y.~Pan, and Q.~Zhu, ``Decision-dominant strategic defense against
  lateral movement for {5G} zero-trust multi-domain networks,'' in
  \emph{Network Security Empowered by Artificial Intelligence}.\hskip 1em plus
  0.5em minus 0.4em\relax Springer, 2024, pp. 25--76.

\bibitem{li2023price}
T.~Li and Q.~Zhu, ``On the price of transparency: A comparison between overt
  persuasion and covert signaling,'' in \emph{2023 62nd IEEE Conference on
  Decision and Control (CDC)}.\hskip 1em plus 0.5em minus 0.4em\relax IEEE,
  2023, pp. 4267--4272.

\bibitem{massicot2019public}
O.~Massicot and C.~Langbort, ``Public signals and persuasion for road network
  congestion games under vagaries,'' \emph{IFAC-PapersOnLine}, vol.~51, no.~34,
  pp. 124--130, 2019.

\bibitem{aid2023signalling}
R.~Aid, A.~Kowli, and A.~A. Kulkarni, ``Signalling for electricity demand
  response: When is truth telling optimal?'' \emph{arXiv preprint
  arXiv:2302.12770}, 2023.

\bibitem{liu2022eproach}
S.~Liu and Q.~Zhu, ``{EPROACH}: A population vaccination game for strategic
  information design to enable responsible {COVID} reopening,'' in \emph{2022
  American Control Conference (ACC)}.\hskip 1em plus 0.5em minus 0.4em\relax
  IEEE, 2022, pp. 568--573.

\bibitem{zhu2010price}
Q.~Zhu and T.~Ba{\c{s}}ar, ``Price of anarchy and price of information in
  {N-person} linear-quadratic differential games,'' in \emph{Proceedings of the
  2010 American Control Conference}.\hskip 1em plus 0.5em minus 0.4em\relax
  IEEE, 2010, pp. 762--767.

\bibitem{zhang2023stochastic}
T.~Zhang and Q.~Zhu, ``Stochastic game with interactive information
  acquisition: A fixed-point alignment principle,'' in \emph{2023 59th Annual
  Allerton Conference on Communication, Control, and Computing
  (Allerton)}.\hskip 1em plus 0.5em minus 0.4em\relax IEEE, 2023, pp. 1--8.

\bibitem{liu2023information}
S.~Liu and Q.~Zhu, ``Information manipulation in partially observable {Markov}
  decision processes,'' \emph{arXiv preprint arXiv:2312.07862}, 2023.

\bibitem{velicheti2023strategic}
R.~K. Velicheti, M.~Bastopcu, and T.~Ba{\c{s}}ar, ``Strategic information
  design in quadratic multidimensional persuasion games with two senders,'' in
  \emph{2023 American Control Conference (ACC)}.\hskip 1em plus 0.5em minus
  0.4em\relax IEEE, 2023, pp. 1716--1722.

\bibitem{VBEB24}
R.~Velicheti, M.~Bastopcu, R.~Etesami, and T.~Ba\c{s}ar, ``Learning how to
  strategically disclose information,'' in \emph{Proc. 2024 American Control
  Conference (ACC 2024)}, Toronto, Canada, July 2024, pp. 1602--1607.

\bibitem{zhu2023doctrine}
Q.~Zhu, ``The doctrine of cyber effect: An ethics framework for defensive cyber
  deception,'' \emph{arXiv preprint arXiv:2302.13362}, 2023.

\bibitem{chen2019control}
J.~Chen and Q.~Zhu, ``Control of multilayer mobile autonomous systems in
  adversarial environments: A games-in-games approach,'' \emph{IEEE
  Transactions on Control of Network Systems}, vol.~7, no.~3, pp. 1056--1068,
  2019.

\bibitem{sandholm2010population}
W.~H. Sandholm, \emph{Population Games and Evolutionary Dynamics}.\hskip 1em
  plus 0.5em minus 0.4em\relax MIT press, 2010.

\bibitem{sandholm2020evolutionary}
W.~H. Sandholm, ``Evolutionary game theory,'' \emph{Complex social and
  behavioral systems: game theory and agent-based models}, pp. 573--608, 2020.

\bibitem{SBB24}
S.~Sanjari, S.~Bose, and T.~Ba\c{s}ar, ``Incentive designs for {Stackelberg}
  games with a large number of followers and their mean-field limits,''
  \emph{Dynamic Games and Applications}, 2024, online 23 May 2024.

\bibitem{farooq2017secure}
M.~J. Farooq and Q.~Zhu, ``Secure and reconfigurable network design for
  critical information dissemination in the {Internet of Battlefield Things
  (IoBT)},'' in \emph{2017 15th International Symposium on Modeling and
  Optimization in Mobile, Ad Hoc, and Wireless Networks (WiOpt)}.\hskip 1em
  plus 0.5em minus 0.4em\relax IEEE, 2017, pp. 1--8.

\bibitem{farooq2018secure}
M.~J. Farooq and Q.~Zhu, ``On the secure and reconfigurable multi-layer network
  design for critical information dissemination in the internet of battlefield
  things {(IoBT)},'' \emph{IEEE Transactions on Wireless Communications},
  vol.~17, no.~4, pp. 2618--2632, 2018.

\bibitem{ameli2022departure}
M.~Ameli, M.~S.~S. Faradonbeh, J.-P. Lebacque, H.~Abouee-Mehrizi, and
  L.~Leclercq, ``Departure time choice models in urban transportation systems
  based on mean field games,'' \emph{Transportation Science}, vol.~56, no.~6,
  pp. 1483--1504, 2022.

\bibitem{huang2021dynamic}
K.~Huang, X.~Chen, X.~Di, and Q.~Du, ``Dynamic driving and routing games for
  autonomous vehicles on networks: A mean field game approach,''
  \emph{Transportation Research Part C: Emerging Technologies}, vol. 128, p.
  103189, 2021.

\bibitem{zhu2013multi}
Q.~Zhu and T.~Ba{\c{s}}ar, ``Multi-resolution large population stochastic
  differential games and their application to demand response management in the
  smart grid,'' \emph{Dynamic Games and Applications}, vol.~3, pp. 68--88,
  2013.

\bibitem{couillet2012electrical}
R.~Couillet, S.~M. Perlaza, H.~Tembine, and M.~Debbah, ``Electrical vehicles in
  the smart grid: A mean field game analysis,'' \emph{IEEE Journal on Selected
  Areas in Communications}, vol.~30, no.~6, pp. 1086--1096, 2012.

\bibitem{dai2024incentive}
X.~Dai, W.~Xu, Y.~Qi, and M.~Jordan, ``Incentive-aware recommender systems in
  two-sided markets,'' \emph{ACM Transactions on Recommender Systems}, vol.~2,
  no.~4, pp. 1--38, 2024.

\bibitem{scheid2024incentivized}
A.~Scheid, D.~Tiapkin, E.~Boursier, A.~Capitaine, E.~M.~E. Mhamdi,
  {\'E}.~Moulines, M.~I. Jordan, and A.~Durmus, ``Incentivized learning in
  principal-agent bandit games,'' \emph{arXiv preprint arXiv:2403.03811}, 2024.

\bibitem{laud2004theory}
A.~D. Laud, ``Theory and application of reward shaping in reinforcement
  learning,'' Ph.D. dissertation, University of Illinois at Urbana-Champaign,
  2004.

\bibitem{zhu2020cross}
Q.~Zhu and Z.~Xu, \emph{Cross-Layer Design for Secure and Resilient
  Cyber-Physical Systems}.\hskip 1em plus 0.5em minus 0.4em\relax Springer,
  2020.

\bibitem{rieger2019industrial}
C.~Rieger, I.~Ray, Q.~Zhu, and M.~Haney, \emph{Industrial Control Systems
  Security and Resiliency}.\hskip 1em plus 0.5em minus 0.4em\relax Springer,
  2019.

\bibitem{rieger2013resilient}
C.~G. Rieger, K.~L. Moore, and T.~L. Baldwin, ``Resilient control systems: A
  multi-agent dynamic systems perspective,'' in \emph{IEEE International
  Conference on Electro-Information Technology, EIT 2013}.\hskip 1em plus 0.5em
  minus 0.4em\relax IEEE, 2013, pp. 1--16.

\bibitem{zhu2021control}
Q.~Zhu, ``Control challenges,'' \emph{Resilient Control Architectures and Power
  Systems}, pp. 215--229, 2021.

\bibitem{AB11}
T.~Alpcan and T.~Ba{\c{s}}ar, \emph{Network Security: A Decision and Game
  Theoretic Approach}.\hskip 1em plus 0.5em minus 0.4em\relax Cambridge
  University Press, 2011.

\bibitem{amini2024control}
A.~Amini, Y.~E. Bayiz, and U.~Topcu, ``Control of misinformation with safety
  and engagement guarantees,'' in \emph{2024 American Control Conference
  (ACC)}.\hskip 1em plus 0.5em minus 0.4em\relax IEEE, 2024, pp. 151--158.

\bibitem{chen2019dynamic}
J.~Chen, C.~Touati, and Q.~Zhu, ``A dynamic game approach to strategic design
  of secure and resilient infrastructure network,'' \emph{IEEE Transactions on
  Information Forensics and Security}, vol.~15, pp. 462--474, 2019.

\bibitem{chen2016interdependent}
J.~Chen and Q.~Zhu, ``Interdependent network formation games with an
  application to critical infrastructures,'' in \emph{2016 American Control
  Conference (ACC)}.\hskip 1em plus 0.5em minus 0.4em\relax IEEE, 2016, pp.
  2870--2875.

\bibitem{huang2017large}
L.~Huang, J.~Chen, and Q.~Zhu, ``A large-scale {Markov} game approach to
  dynamic protection of interdependent infrastructure networks,'' in
  \emph{International Conference on Decision and Game Theory for
  Security}.\hskip 1em plus 0.5em minus 0.4em\relax Springer, 2017, pp.
  357--376.

\bibitem{huang2018factored}
L.~Huang, J.~Chen, and Q.~Zhu, ``Factored {Markov} game theory for secure
  interdependent infrastructure networks,'' \emph{Game Theory for Security and
  Risk Management: From Theory to Practice}, pp. 99--126, 2018.

\bibitem{EB18}
S.~Etesami and T.~Ba\c{s}ar, ``Network games,'' in \emph{Handbook of Dynamic
  Game Theory}, T.~Ba\c{s}ar and G.~Zaccour, Eds.\hskip 1em plus 0.5em minus
  0.4em\relax Springer International Publishing, 2018, vol.~1, ch.~12, pp.
  547--593.

\bibitem{zhao2022multi}
Y.~Zhao, C.~Rieger, and Q.~Zhu, ``Multi-agent learning for resilient
  distributed control systems,'' \emph{arXiv preprint arXiv:2208.05060}, 2022.

\bibitem{TB83}
T.~Ba\c{s}ar, ``Performance bounds for hierarchical systems under partial
  dynamic information,'' \emph{Journal of Optimization Theory and
  Applications}, vol.~39, no.~1, pp. 67--87, January 1983.

\bibitem{TB89}
T.~Ba\c{s}ar, ``Stochastic incentive problems with partial dynamic information
  and multiple levels of hierarchy,'' \emph{European J. Political Economy
  (special issue on ``Economic Design'')}, vol.~V, pp. 203--217, 1989.

\bibitem{TB89b}
T.~Ba\c{s}ar, ``Time consistency and robustness of equilibria in noncooperative
  dynamic games,'' in \emph{Dynamic Policy Games in Economics}, F.~V. der Ploeg
  and A.~de~Zeeuw, Eds.\hskip 1em plus 0.5em minus 0.4em\relax North Holland,
  1989, pp. 9--54.

\end{thebibliography}

\end{document}